\documentclass[floats,floatfix,showpacs,amssymb,prd,onecolumn,superscriptaddress,nofootinbib]{revtex4-2}

\usepackage{graphicx}
\usepackage{dcolumn}
\usepackage{bm}
\usepackage{amsmath}
\usepackage{color}

\definecolor{red}{rgb}{1,0,0}
\definecolor{darkred}{rgb}{0.6,0,0}
\definecolor{darkgreen}{rgb}{0.992447,0.623778,0.034597}
\definecolor{ppink}{rgb}{1,0.4,0.4}
\definecolor{bblue}{rgb}{0.284602,0.317763,0.963947}
\definecolor{purple}{rgb}{0.5 ,0, 0.7}
\def\beq{\begin{equation}}
\def\eeq{\end{equation}}

\usepackage[final]{hyperref} 
\hypersetup{
    colorlinks=true,       
    linkcolor=blue,        
    citecolor=blue,        
    filecolor=magenta,     
    urlcolor=blue         
}

\usepackage{xcolor}
\usepackage[normalem]{ulem}
\usepackage{here}

\newcommand{\dd}{{\text d}}

\begin{document}

 \begin{flushright}
DESY-24-219
\end{flushright}

\title{Gravitational Waves from Particles Produced from Bubble Collisions\\ in First-Order Phase Transitions}

\author{Keisuke Inomata}
\email{kinomat1@jhu.edu}
\affiliation{William H.\ Miller III Department of Physics \& Astronomy, Johns Hopkins University, 3400 N.\ Charles St., Baltimore, MD 21218, USA}

\author{Marc Kamionkowski}
\email{kamion@jhu.edu}
\affiliation{William H.\ Miller III Department of Physics \& Astronomy, Johns Hopkins University, 3400 N.\ Charles St., Baltimore, MD 21218, USA}

\author{Kentaro Kasai}
\email{kkasai@icrr.u-tokyo.ac.jp}
\affiliation{ICRR, University of Tokyo, Kashiwa, 277-8582, Japan}

\author{Bibhushan Shakya}
\email{bibhushan.shakya@desy.de}
\affiliation{Deutsches Elektronen-Synchrotron DESY, Notkestr.\,85, 22607 Hamburg, Germany}

\date{\today}

\begin{abstract} 
We discuss a new source of gravitational waves (GWs) from first-order phase transitions. The collisions of bubbles of the new phase can efficiently produce particles that couple to the background field undergoing the transition, thereby transferring a significant fraction of the released vacuum energy into a distribution of inhomogeneous and dynamic particle populations that persist long after the bubbles have disappeared. We study the GWs produced by such particle distributions, showing that GWs arise from the quadrupolar anisotropy in the radiation emitted from the bubble collisions, and present a semi-analytical calculation of the two-point correlation function for the associated energy distributions. We find that this new contribution can qualitatively modify the overall GW signal from such phase transitions, creating a distinct shift in the spectral slope at low frequencies that could be observed by future GW experiments. It is therefore important to take this new contribution into account for any transition where the background field has significant self-coupling or couplings to other fields that could lead to efficient particle production at bubble collision. 
\end{abstract}

\maketitle

\tableofcontents

\section{Introduction}
Gravitational waves (GWs) from first-order phase transitions (FOPTs) in the early Universe~\cite{Witten:1984rs,Hogan:1986qda,Turner:1990rc,Krauss:1991qu,Kosowsky:1991ua,Kosowsky:1992rz,Kosowsky:1992vn,Kamionkowski:1993fg} have been widely studied as well-motivated and promising targets for current and upcoming GW detectors~\cite{LISA:2017pwj,Harry:2006fi,Kawamura:2011zz,Punturo:2010zz,Sesana:2019vho,Reitze:2019iox}. Although the electroweak phase transition and the QCD phase transition within the Standard Model (SM) are known to not be first-order, the detection of a cosmological FOPT would be a smoking gun for physics beyond the Standard Model (BSM). Indeed, there are numerous BSM frameworks that predict FOPTs in dark sectors~\cite{Schwaller:2015tja,Caprini:2018mtu,Jaeckel:2016jlh,Baldes:2017rcu, Hall:2019rld, Tsumura:2017knk,Croon:2018erz,Baldes:2018emh,Prokopec:2018tnq,Breitbach:2018ddu} that can produce observable GW signals. 

FOPTs, which proceed through the nucleation, expansion, and percolation of bubbles of the new phase, are known to produce GWs in various ways (see e.g.\cite{Caprini:2015zlo,Caprini:2018mtu,Caprini:2019egz,LISACosmologyWorkingGroup:2022jok, Athron:2023xlk} for some reviews). In scenarios where bubble walls carry the dominant fraction of the energy released in the transition, GWs are produced from the scalar field energy densities in the bubble walls when the walls collide~\cite{Kosowsky:1991ua,Kosowsky:1992rz,Kosowsky:1992vn,Kamionkowski:1993fg,Huber:2008hg,Bodeker:2009qy,Jinno:2016vai,Jinno:2017fby,Konstandin:2017sat,Cutting:2018tjt,Cutting:2020nla}. If the energy is instead primarily transferred to the plasma surrounding the bubble walls during the bubble expansion stage, GWs are instead produced from sound waves~\cite{Hindmarsh:2013xza,Hindmarsh:2015qta,Hindmarsh:2017gnf,Cutting:2019zws,Hindmarsh:2016lnk,Hindmarsh:2019phv,Guo:2024gmu} and turbulence~\cite{Kamionkowski:1993fg,Kosowsky:2001xp,Gogoberidze:2007an,Caprini:2009yp,Brandenburg:2017neh,Cutting:2019zws,RoperPol:2019wvy,Dahl:2021wyk,Auclair:2022jod} in the plasma, or through dynamic distributions of feebly interacting particles~\cite{Jinno:2022fom} if the particles surrounding the bubbles do not have significant interactions to warrant a coherent fluid(plasma)-like behavior.   

In this paper, we study a new source of GWs from FOPTs: particles produced from the collisions of bubbles. The production of quanta of fields that couple to the background field undergoing the phase transition during the bubble collision process has been studied in several papers \cite{Watkins:1991zt,Konstandin:2011ds,Falkowski:2012fb,Mansour:2023fwj,Shakya:2023kjf} (and subsequently found BSM applications for dark matter \cite{Falkowski:2012fb,Freese:2023fcr, Giudice:2024tcp} and leptogenesis \cite{Cataldi:2024pgt}), which established this phenomenon to be an efficient process that can transfer a significant fraction of the energy released during the phase transition into the produced particle distribution. This energy component is generally ignored in studies of GWs from FOPTs. In the simplest approach to calculate GWs from bubble collisions, termed the ``envelope approximation" \cite{Kosowsky:1992vn,Kamionkowski:1993fg,Huber:2008hg,Jinno:2016vai}, the energy in the collided regions is assumed to dissipate very quickly, and thus neglected for the purposes of computing the GW signal. In the so-called ``bulk flow model" \cite{Jinno:2017fby,Konstandin:2017sat}, this simplistic approach is extended by propagating the shells of shear-stress forward after the bubbles collide. While the bulk flow model is applicable to some degree to the particle production process we have in mind, it is an extremely simple ansatz aimed at capturing the basic idea that some remnant of the energy contained in the bubble walls should continue to contribute after collision. 

In this paper, we model the energy density in particles produced from bubble collisions more realistically to better capture the underlying physics of the process. We find that GWs are produced due to the quadrupolar anisotropy in the spectrum of the particles produced from bubble collisions. We derive a semi-analytic formula for the GW spectrum sourced by such particle distributions. This GW component is smaller in amplitude than the standard GW signal from the scalar field configuration of colliding bubbles, but falls only linearly with frequency in the infrared (IR), and therefore can qualitatively modify the IR component of the overall GW signal from FOPTs. Our results are thus important for, and applicable to, any FOPT where the background field has significant ($\mathcal{O}(1)$) self-coupling or couplings to other fields, which will result in efficient particle production at bubble collision, and must be taken into account to compute the full GW signal.  

This paper is organized as follows. 
In Sec.~\Ref{framework}, we discuss the framework of our study, and the physics of particle production from bubble collisions. Sec.~\Ref{Calculation of Stress-Energy Tensor}
presents our calculation of the two-point correlation function of the stress-energy tensor for the distribution of such particles. The subsequent GW spectrum is presented in Sec.~\Ref{sec : GW spectrum}. Sec.~\Ref{sec : Discussions and conculsions} is devoted to a broader discussion of our results and conclusions.


\section{Framework}
\label{framework}

We are interested in FOPT scenarios where particles produced at bubble collisions can account for a significant fraction of the energy released during the phase transition. This is necessary if the particle distribution is to produce an observable effect on the overall GW signal arising from the FOPT. There are two main pre-requisites for realizing such scenarios.
\begin{itemize}
    \item The bubble walls should retain a significant fraction of the energy released over the duration of the phase transition as the bubbles expand, i.e. achieve the so-called runaway behavior.
    \item Bubble collisions should result in efficient particle production, such that a large fraction of the energy carried by the walls gets converted to particles.
    \end{itemize}

In this section, we briefly discuss the underlying physics that enables these conditions to be satisfied.

\subsection{Setup}

We adopt a model-independent approach for generality, basing our discussions on phenomenological parameters that can be calculated in a given model.  For specificity, consider a FOPT involving a background scalar field $\phi$ transitioning from a metastable vacuum characterized by vanishing vacuum expectation value (VEV) $\langle \phi \rangle=0$ to a stable vacuum with $\langle \phi \rangle=v_\phi$.  We can parameterize the difference in the potential energies of the two vacua as 
\beq
\Delta V \equiv V_{\langle \phi \rangle=0}-V_{\langle \phi \rangle=v_\phi}=c_V\,v_\phi^4\,.
\label{deltaV}
\eeq
The following phenomenological FOPT parameters will be relevant for our analysis:

\begin{itemize}
\item $\alpha\equiv\frac{\rho(\text{vacuum})}{\rho(\text{radiation})}=
\frac{\Delta V}{\rho_{0}}$: the strength of the phase transition, where $\rho_0$ is the energy density in the radiation bath at the time of the transition.

\item $\beta$: (inverse) duration of the phase transition. This is generally written in units of the Hubble parameter $H$, as $\beta/H$. This also determines the typical bubble size at collision, $R_*\approx (8\pi)^{1/3}{\beta}^{-1}$.

\item $R_0$: initial bubble radius at nucleation.

\item $\gamma=1/\sqrt{1-v^2}$: the Lorentz boost factor of the bubble wall, where $v$  is the bubble wall velocity. In this paper, we are interested in the ultra-relativistic regime $v\approx 1,~\gamma\gg 1$. In the runaway regime, the boost factor is known to grow linearly with the size of the bubble, $\gamma\sim R/R_0$.

\item $l_w$: bubble wall thickness (in the cosmic frame); it Lorentz contracts as $l_{w_0}\to l_w=l_{w_0}/\gamma$, where  $l_{w_0}\sim\mathcal{O}(v_\phi^{-1})$ is the initial thickness at nucleation.

\item $T$: the temperature of the thermal bath during the phase transition. 
\end{itemize}

 In general, an expanding bubble encounters friction due to interactions with the surrounding plasma, which take away energy from the bubble and slow it down. 
 For the bubble walls to retain a significant fraction of the energy released in the phase transition, such friction terms must be negligible or absent. 
 This can occur in several scenarios (see \cite{Giudice:2024tcp} for a detailed discussion), such as supercooled phase transitions \cite{Konstandin:2011dr,vonHarling:2017yew,Ellis:2018mja,Baratella:2018pxi,DelleRose:2019pgi,Fujikura:2019oyi,Ellis:2019oqb,Brdar:2019qut,Baldes:2020kam}, transitions that occur via quantum tunneling in a cold sector, or transitions where splitting radiation involving gauge bosons \cite{Bodeker:2017cim,Hoche:2020ysm, Azatov:2020ufh,Gouttenoire:2021kjv,Ai:2023suz,Azatov:2023xem,Long:2024sqg}, which is known to produce a significant friction effect that is proportional to the wall boost factor, is absent. In such cases, the bubble walls achieve the so-called runaway behavior, retaining a significant fraction of the energy released over the course of the phase transition in the form of kinetic and gradient energy. We now turn to the phenomenon of particle production from the collisions of such runaway bubble walls.

\subsection{Particle production from bubble collisions}
\label{sec:particle_prod} 

Consider particle production from two planar walls with the same thickness colliding in the $z$-direction. Here we provide a brief outline of the formalism to calculate particle production from such collisions; the interested reader is referred to \cite{Watkins:1991zt,Falkowski:2012fb,Mansour:2023fwj,Shakya:2023kjf,Giudice:2024tcp} for greater details. In this formalism, particle production is evaluated by calculating the imaginary part of the effective action of the background field configuration, which is Fourier decomposed into modes of definite four-momenta $k^2=\omega^2-k_z^2>0$ (where $\omega$ and $k_z$ are the frequency and momentum obtained from the Fourier transform), which are interpreted as off-shell propagating field quanta $\phi^*_k$ of the background field with effective mass $m^2=k^2$ that can decay into particles. The number of particles produced per unit area of bubble collision (assuming planar bubble walls) can be written as 
\cite{Watkins:1991zt,Falkowski:2012fb}
 \beq
\frac{N}{A}=\frac{1}{2 \pi^2}\int_{k_{\text{min}}^2}^{k_{\text{max}}^2} d k^2\,f(k^2) \,\text{Im} [\tilde{\Gamma}^{(2)}(k^2)] \, .
\label{number}
\eeq
The lower limit of the integral is determined by either the mass of the particle species being produced, i.e.\,$k_{\text{min}}=2m$ for pair-production, or by the inverse size of the bubble, $k_{\text{min}}=(2R_*)^{-1}$. The upper limit is given by the maximum energy available in the collision process, $k_{\text{max}}=2/l_w=2\gamma/l_{w_0}$.

The $f(k^2)$ term represents the efficiency factor for particle production at a given scale $p$, and depends on the nature of the collision and the subsequent dynamics of the background field. The component that accounts for the collision of ultrarelativistic bubbles takes a universal form 
\cite{Watkins:1991zt,Falkowski:2012fb,Mansour:2023fwj,Shakya:2023kjf}
\beq
f(k^2)=\frac{16 v_{\phi}^2}{k^4}\, \text{Log}\left[\frac{2(1/l_w)^2-k^2+2(1/l_w)\sqrt{(1/l_w)^2-k^2}}{k^2}\right]\,.
\label{eq:felastic}
\eeq
The imaginary part of the 2-point 1PI Green function $\Gamma^{(2)}$ can be calculated using the optical theorem \cite{Watkins:1991zt,Falkowski:2012fb}, and depends on the nature of the interaction. The $\phi$ particles themselves are produced at bubble collision due to self-interactions of the form $\frac{\lambda_\phi}{4!}\phi^4$; this gives rise to 2- and 3-body decay channels with
\beq
\text{Im} [\tilde{\Gamma}^{(2)}(k^2)]_{\phi^*_k\to\phi\phi}\approx\frac{\lambda_\phi^2 \,v_\phi^2}{8\pi},~~~~~\text{Im} [\tilde{\Gamma}^{(2)}(k^2)]_{\phi^*_k\to3\phi}\approx\frac{\lambda_\phi^2 \,k^2}{3072\,\pi^3}.
\label{23scalar}
\eeq
Similarly, a Yukawa coupling to a fermion, $y_f \phi  f\bar{f}$, gives
\beq
\text{Im} [\tilde{\Gamma}^{(2)}(k^2)]_{\phi^*_p\to f\bar{f}}\approx\frac{y_f^2}{8\pi} k^2\,. 
\label{fermion}
\eeq
The calculation of gauge boson production is more complicated because the formalism is not gauge-invariant \cite{Giudice:2024tcp}, but is similar to the pair production of scalars for ultrarelativistic bubble collisions. 

The energy density in particles per unit area is 
 \beq
\frac{E}{A}=\frac{1}{2\pi^2}\int_{k_{\text{min}}^2}^{k_{\text{max}}^2} d k^2\,k\,f(k^2) \,\text{Im} [\tilde{\Gamma}^{(2)}(k^2)] \, .
\label{energy}
\eeq
Using the above relations and definitions, we can estimate the fraction of the vacuum energy released from a single bubble that goes into particles as
\beq
\kappa\equiv\frac{\frac{1}{2}(E/A)\cdot 4\pi R_*^2}{\frac{4}{3}\pi R_*^3\cdot \Delta V}= \frac{3}{2}\frac{E/A}{R_* \Delta V},
\label{eq : kappa}
\eeq
The factor of $1/2$ in the numerator accounts for the fact that particle production at collision comes from the energy released from two bubbles. 

We can perform a rough estimate for this ratio for the case of fermion production (Eq.\,\eqref{fermion}) using Eqs.\,\eqref{deltaV} and \eqref{eq : kappa}, $\gamma\approx {R_*/R_0}$, and $R_0\sim l_{w0} \sim v_\phi^{-1}$; this yields 
\beq
\kappa\sim \frac{6\, y_f^2\,\text{Log[...]}}{\pi^3\, c_V}\,,
\eeq
where Log[...] is the logarithmic factor in Eq.\,\eqref{eq:felastic}, which takes a value between 6 and 60 over regions of parameter space of interest \cite{Giudice:2024tcp}. Thus, $\kappa\sim 1$ is possible for $y_f\sim \mathcal{O}(1)$ and $c_v \lesssim 1$, i.e.\,a significant fraction of the energy released during the phase transition can go into the produced particles.\footnote{Note that, in this case, particle production would backreact on the bubble collision dynamics, which is not accounted for in the above formalism.}

From Eq.\,\eqref{energy}, we can also obtain the energy distribution in the produced particles; assuming that the particles are relativistic, $k\approx E$, and ignoring the approximately constant logarithmic factor in $f(k^2)$, we have 
\beq
\frac{d(E/A)}{dE} \sim \frac{v_\phi^2}{E^2} \,\text{Im} [\tilde{\Gamma}^{(2)}(k^2)]\,.
\eeq
For the fermion (or 3-body scalar) case, $\text{Im} [\tilde{\Gamma}^{(2)}(k^2)]\sim k^2 \sim E^2$, hence $\frac{d(E/A)}{dE}$ is constant, i.e.\,most of the energy is concentrated at the highest values. In contrast, for 2-body decays into scalars and vectors, $\text{Im} [\tilde{\Gamma}^{(2)}(k^2)]\sim v_\phi^2$, hence $\frac{d(E/A)}{dE}\sim 1/E^2$, and the spectrum is instead peaked at lower energies, but a significant fraction of the particles nevertheless remain relativistic. Therefore, it is reasonable to assume that a significant fraction of the energy released during the phase transition can be converted into relativistic particles during bubble collisions.  

These particles can lose energy via scatterings with other particles in the bath.
Naively, one would expect such scatterings to be very efficient in the presence of $\mathcal{O}(1)$ couplings between the produced particle and the $\phi$ field, as would be required for significant particle production through bubble collisions. However, there are two mitigating factors: (i) since these particles are highly boosted, such scatterings have large center-of-mass energies, which suppresses the scattering cross-section, and (ii) multiple scatterings are required for such an energetic particle to lose most of its energy to the bath.  We can estimate the condition for such scatterings to dissipate an $\mathcal{O}(1)$ fraction of the particle energy within a Hubble time as \cite{Baldes:2022oev, Giudice:2024tcp}
\beq
\frac{n_\phi\,\sigma_{\phi f\to\phi f}\, v}{H} \frac{\delta p_{f}}{p_{f}}>1 ~~~\Rightarrow~~~\frac{n_\phi}{T^3}\,y_f^2> 10^3\,\frac{k_\text{max}^2}{T\,M_{Pl}}\,,
\eeq
where we have used various expressions and relations from \cite{Baldes:2022oev}, and specialized to the case of fermion production. From this, we see that even with  $y_f\sim\mathcal{O}(1)$, such scatterings can be inefficient if $k_\text{max}^2\gtrsim T\,M_{Pl}$. It is also worth noting that such scatterings are subdominant for supercooled transitions, where the pre-existing thermal bath gets diluted by a brief period of inflationary expansion. 

In addition, we will assume that these particles are stable over a Hubble time. This is generally true if these are dark sector particles that decay into SM final states through tiny portal couplings, and is further aided by the fact that these particles are extremely boosted, hence their decays are time-dilated. However, we also note that they cannot be extremely long-lived, otherwise they can overclose the Universe, and therefore must decay into SM states, likely before Big Bang Nucleosynthesis.      

Based on the above discussions, we will restrict our attention to the following configuration as a viable framework for the remainder of the paper:
\begin{itemize}
\item Bubble collisions transfer a significant fraction of the vacuum energy released during the phase transition into particles produced from the collisions. 
\item The produced particles are relativistic, i.e.\,their masses are negligible compared to their energies, and they propagate out of the bubble collision sites at the speed of light. 
\item The particles free-stream without interacting or decaying, and consequently the energy stored in the particle ensemble is retained without loss, over a Hubble time. 
\end{itemize}

\section{Calculation of Stress-Energy Tensor}
\label{Calculation of Stress-Energy Tensor}

In this section, we present our calculation of the stress-energy tensor for the distribution of particles produced from bubble collisions, and its two-point correlation function. These will be used to calculate the GWs generated by the particle distribution in the next section. 

\subsection{Modeling the energy distribution of particles from bubble collisions}

In the cosmic frame, bubbles generally have unequal sizes, and consequently bubble walls have unequal velocities and energies at collision. Consider the collision of two bubble wall segments with velocities $\bm{v}_a$, $\bm{v}_b$ in the cosmic frame. We define $f_{\rm{coll}}$, the momentum distribution function of the emitted particles from this collision, as
\begin{equation}
    \int
    \frac{{\rm{d}}^3\bm{p}}
    {(2\pi)^3}
    p f_{\rm{coll}}
    \left(
    \bm{v}_a, \bm{v}_b
    ,\bm{p}
    \right)
    = 
    \rho_p(v_a, v_b),
    \label{eq : definition of f coll}
\end{equation}
where $\rho_p$ is the total energy density of the emitted particles. 
Recall that we assume that the emitted particles are relativistic. 
Note that $v_{a(b)} = |\bm v_{a(b)}|$, and we use the same notation for any 3-dimensional vector in the following unless otherwise stated.

Likewise, we define the energy flux of the emitted particles in a given direction $\hat{\bm{p}}(\equiv \bm p/p)$ as 
\begin{align}
    \rho_{\Omega}(\bm{v}_a,\bm{v}_b,\hat{\bm{p}})
    \equiv 
    \int
    \frac{{\rm{d}}p}{(2\pi)^3}
    p^3
    f_{\rm{coll}}(\bm{v}_a,\bm{v}_b,\bm{p}).
    \label{eq : definition of rho Omega}
\end{align}
Due to the asymmetry in the energy densities and velocities of the two colliding bubble wall fragments, the momentum distribution of the produced particles has an angular anisotropy.  

\begin{figure}[t]
    \centering
\includegraphics[width=0.75\linewidth]{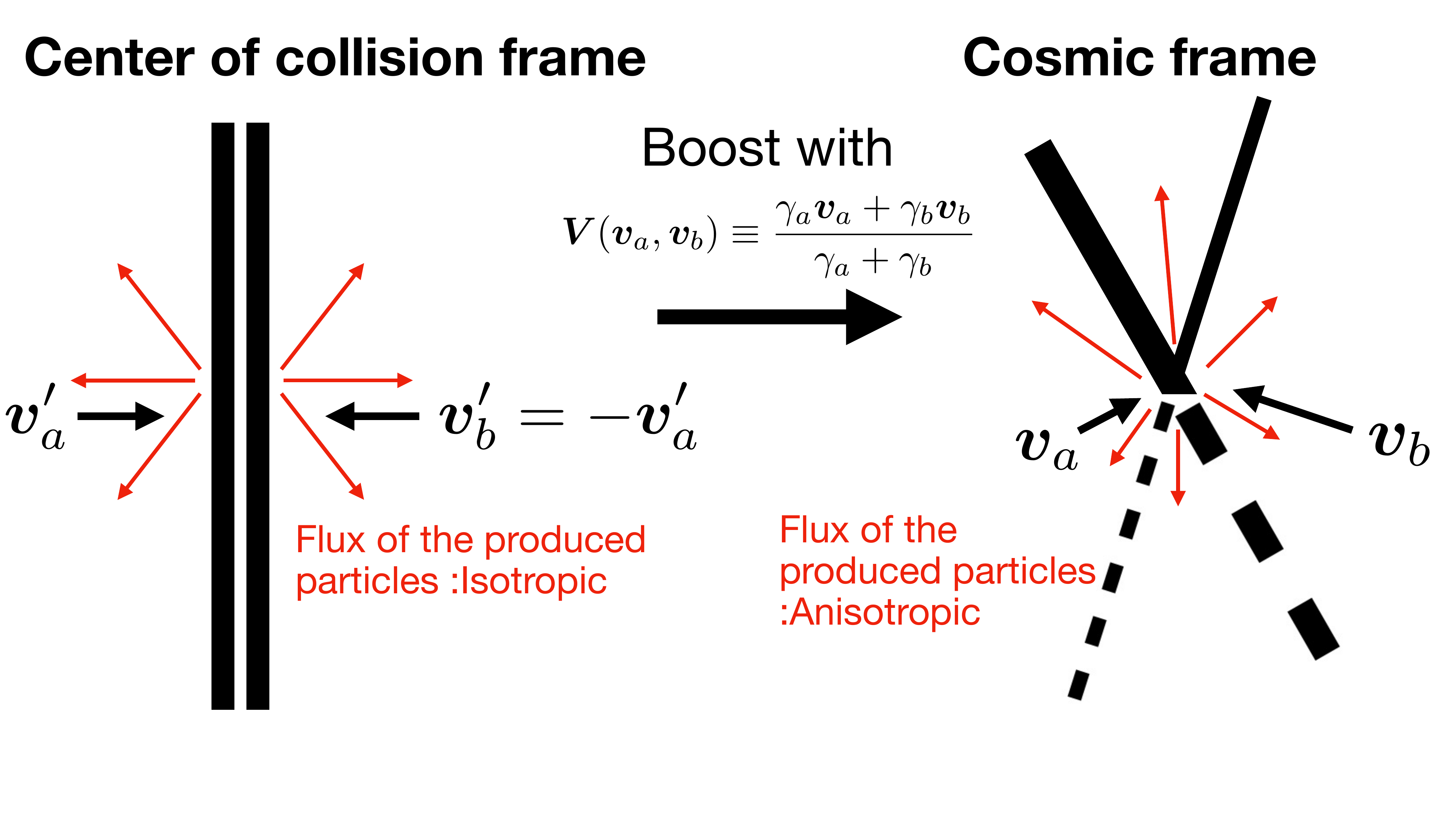}
        \caption{ 
        A schematic illustration of the spectrum of particles produced from bubble collision. The distribution is isotropic in the ``center of collision" frame, where the two bubbles have the same size and the colliding segments have equal and opposite velocities. This distribution can then be boosted to the cosmic frame to obtain the anisotropic distribution of particles that will produce GWs. 
        }
\label{fig:boost}
\end{figure}

To calculate this anisotropy, we can first consider the bubble collision in the ``center of collision frame", characterized by  $\bm{v}_a' = -\bm{v}_b'$. Note that there always exists a rest frame where this is true: one can first go to a rest frame where the two bubbles nucleate simultaneously, so that the two bubbles walls have the same size and energy at collision (since the bubbles are of equal size), i.e. $v_a' = v_b'$, then perform additional boosts so that the two velocities are head-on. In this frame, since the collision is symmetric, we assume that the off-shell excitations $\phi^*_k$ produced from the collision are at rest, and the particles produced from their decay are isotropic. Note that these excitations follow the dispersion relation $k^2=\omega^2-k_z^2$, hence carry some physical momenta $k_z$ even in this frame. However, recall from the previous section that most of the energy in the produced particles is concentrated towards the highest values, $k\approx k_\text{max}$. For these, $k_z^2/ k^2\lesssim 1$, hence it should be reasonable to approximate the distribution of particles produced from their decays as isotropic, up to $\mathcal{O}(1)$ corrections. This is in stark difference with the bulk flow model, which assumes that all produced particles propagate along $\bm{v}_a'$ or $\bm{v}_b'$ even in this frame. Our formalism therefore better captures the underlying physics compared to the bulk flow model in this regard.

We can now boost this isotropic distribution of produced particles from this ``center of collision" frame to the cosmic frame in order to obtain the particle distribution that will be relevant for the generation of GWs (see Fig.\,\ref{fig:boost} for a schematic illustration). 
We calculate that this boost corresponds to a velocity (see Appendix~\Ref{append:boost} for details) 
\begin{align}
    \bm{V}
    \equiv
    \frac{\gamma_a\bm{v}_a+\gamma_b\bm{v}_b}{\gamma_a+\gamma_b}\,,
    \label{boostvelocity}
\end{align}
where $\gamma_a\equiv 1/\sqrt{1-v_a^2}$, $\gamma_b\equiv 1/\sqrt{1-v_b^2}$.  
After this boost, the particle energy distribution in the cosmic frame can be approximated as (see Appendix~\Ref{append:boost})
\begin{align}
   \rho_{\Omega}(\bm{v}_a,\bm{v}_b,\hat{\bm{p}})
    &\simeq \frac{1}{4\pi}
     \rho_p(v_a, v_b)
    \left(
    1+3\bm{V}\cdot \hat{\bm{p}}+6(\bm{V}\cdot \hat{\bm{p}})^2
    \right).
    \label{boosted_distribution}
\end{align}

Here, $\rho_p$ can be written as a fraction of the energy densities of the two colliding wall segments 
\begin{align}
    \rho_p(v_a, v_b)=\kappa 
    \left(
    \rho_{\rm{wall}}(v_a)+\rho_{\rm{wall}}(v_b)
    \right).
    \label{eq : definition of rho p}
\end{align}
Here we assume that all of the released vacuum energy has been transferred to the bubble walls, so that 
\begin{align}
    \rho_{\rm{wall}} (v)
    =
    \frac{\kappa_\rho}{3} 
    \frac{R (v)}{l_w}
    \Delta V.
    \label{eq : 
    definition of kappa}
\end{align}
Note that $R$, the radius of the bubble, is a function of $v$, and that the two bubbles can have different radii at collision. 

\subsection{Stress-energy tensor}
 \label{sec : concrete expression of the stress tensor}
The spatial components of the stress-energy tensor induced by particles at spacetime $(t,\bm{x})$ are given by~\cite{Weinberg:2008zzc}
\begin{equation}
T_{ij}(t,\bm{x})=
\int\frac{{\rm{d}}^{3}\bm{p}}{(2\pi)^3}
f(t,\bm{x},p,\hat{\bm{p}})
\frac{p_ip_j}{E(\bm{p})}
=\int\frac{{\rm{d}}^{3}\bm{p}}{(2\pi)^3}
f(t,\bm{x},p,\hat{\bm{p}})
p\hat{p}_i\hat{p}_j,
\label{eq : definition of Tij}
\end{equation}
where $f(t,\bm{x},p,\hat{\bm{p}})$ is the momentum distribution function of the particles with momentum $\bm{p}$ at $(t,\bm{x})$.  $E(\bm{p})\simeq p$ is the energy of the relativistic particle with momentum $\bm{p}$, and in the last equality, we assumed that the produced particles are relativistic.  We have also assumed that the emitted particles free-stream with velocity $\bm{p}/E(\bm{p})\simeq \hat{\bm{p}}$ once produced. 
The momentum distribution function can be expressed as
\begin{align}
f(t,\bm{x},p,\hat{\bm{p}})
=
\int{\rm{d}}^3\bm{x}'
\int_{t_{\rm{start}}}^{t}
\frac{{\rm{d}}t'}{\Delta t_{\rm{coll}}}
f_{\rm{coll}}
\left(\bm{v}_a(t',\bm{x}+\bm{x}'),\bm{v}_b(t',\bm{x}+\bm{x}'),\bm{p}
\right)
\xi(t',\bm{x}+\bm{x}')
\delta^{(3)}
\left(
\bm{x}'-\hat{\bm{p}}(t-t')
\right).
\label{eq : f}
\end{align}
Here, we have treated $\bm{v}_a$ and $\bm{v}_b$ as functions of the collision point coordinates, and $\Delta t_{\rm{coll}}$, the typical timescale of the collision process, is taken to be $\Delta t_{\rm{coll}} \approx \gamma^{-1} l_{w0} =\beta R_0 l_{w0}$. 
$t_{\text{start}}$ represents the onset of the phase transition, and the stochastic variable $\xi$ is defined as
\begin{align}
\xi(t',\bm{x}+\bm{x}')\equiv
\begin{cases}
1 &
{\rm{when\;two\;bubble\;walls\;collide\;at\;}}(t',\bm{x}+\bm{x}')
\\
0 &{\rm{otherwise}}
\end{cases}.
\end{align}

Using Eqs.~\eqref{eq : definition of f coll},~\eqref{eq : definition of rho Omega},~\eqref{eq : definition of Tij} and~\eqref{eq : f}, the stress-energy tensor can be written as
\begin{align}
T_{ij}(t,\bm{x})
&= 
(\beta R_0l_{w0})^{-1}
\int{\rm{d}}^3\bm{x'}
\int_{t_{\rm{start}}}^{t}
{\rm{d}}t'
\xi(t',\bm{x}+\bm{x}')
\rho_\Omega
(\bm{v}_a(t',\bm{x}+\bm{x}'),
\bm{v}_b(t',\bm{x}+\bm{x}'), \hat{\bm{x}}')
W
(x',t-t'
)
\hat{\bm{x}}_i'\hat{\bm{x}}_j',
\nonumber \\
&= 
(\beta R_0l_{w0})^{-1}
\int_{t_{\rm{start}}}^{t}{\rm{d}}t'\int{\rm{d}}^3\bm{x'}
\xi(t',\bm{x}+\bm{x}')
\rho_\Omega
(\bm{v}_a(t',\bm{x}+\bm{x}'),
\bm{v}_b(t',\bm{x}+\bm{x}'), \hat{\bm{x}}')
W
(x',t-t'
)
\hat{\bm{x}}_i'\hat{\bm{x}}_j',
\label{eq : deformation of energy momentum tensor 2}
\end{align}
where $\hat{\bm{x}}'= \hat{\bm{p}}$ is the unit vector proportional to $\bm{x}'$.
We have defined the window function $W$ as
\begin{equation}
    W(x',t-t')
    \equiv
\int{\rm{d}}\Omega_{\bm{p}}\delta^{(3)}(\bm{x}'-\hat{\bm{p}}(t-t'))=
    \frac{1}{{x'}^2}
    \delta\left(x'-(t-t')\right).
    \label{eq : definition of W}
\end{equation}
The $\delta\left(x'-(t-t')\right)$ factor picks out the contribution from the collision process at coordinates $(t',\bm{x}+\bm{x'})$, which satisfies $x'=t-t'$. 

\begin{figure}[t]
\centering
\includegraphics[width=0.8\linewidth]{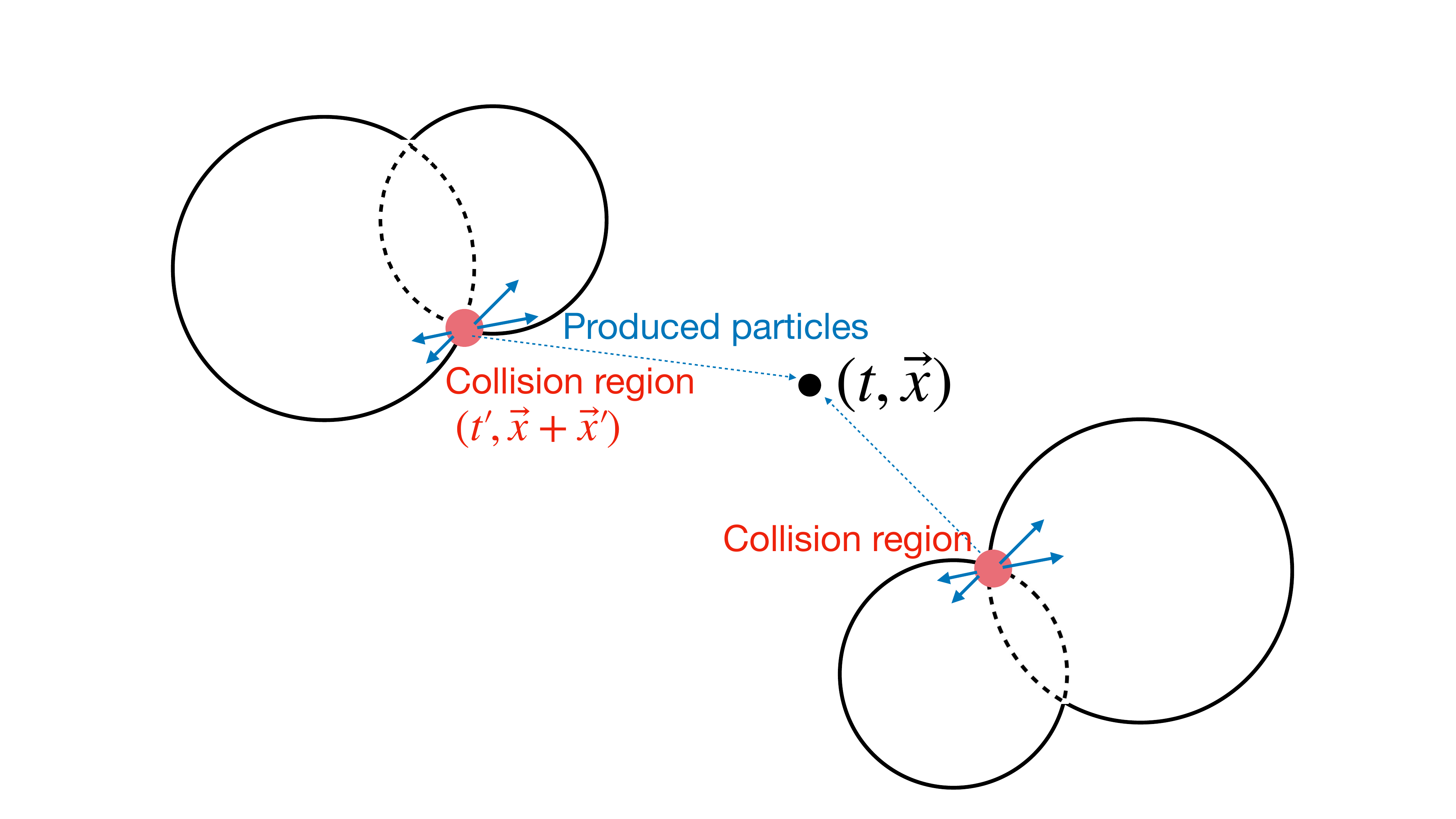}
    \caption{Schematic representation of the contributions to the stress-energy tensor at $(t,\bm x)$.}
    \label{fig : collision regions and the point where we evaluate stress tensor}
\end{figure}

\subsection{Two-point correlation function of transverse-traceless part of the stress-energy tensor}
\label{sec : Power spectrum for transverse-traceless part of the stress tensor}

Next, we calculate the two-point correlation function of the stress-energy tensor, which is needed to calculate the GW spectrum.
Changing the integration variable from $\bm{x}$ to $\bar{\bm{x}}\equiv \bm{x}+\bm{x}'$, and using Eqs.~\eqref{boosted_distribution} and~\eqref{eq : deformation of energy momentum tensor 2}, the Fourier transform of $T_{ij}$ is given by
\begin{align}
T_{ij}(t,\bm{k})
&
=
(\beta R_0l_{w0})^{-1}
\int_{t_{\rm{start}}}^t {\rm{d}}t'
\int {\rm{d}}^3\bm{\bar{x}}\;e^{-i{\bm{k}}\cdot \bm{\bar{x}}}
\xi(t',\bar{\bm{x}})
\int {\rm{d}}^3\bm{x}'
e^{i{\bm{k}}\cdot \bm{x}'}
W
(x',t-t'
)
\rho_{\Omega}(\bm{v}_a(t',\bar{\bm{x}}),\bm{v}_b(t',\bar{\bm{x}}),\hat{\bm{p}})
\hat{\bm{x}}_i'
\hat{\bm{x}}_j'
\nonumber \\
&=
(\beta R_0l_{w0})^{-1}
\left[\int_{t_{\rm{start}}}^t {\rm{d}}t'
\int {\rm{d}}^3\bm{\bar{x}}\;e^{-i{\bm{k}}\cdot \bm{\bar{x}}}
\xi(t',\bar{\bm{x}})
\rho_p(v_a(t',\bar{\bm{x}}),v_b(t',\bar{\bm{x}}))
\int {\rm{d}}^3\bm{x}'
e^{i{\bm{k}}\cdot \bm{x}'}
W
(x',t-t'
)
\hat{\bm{x}}_i'\hat{\bm{x}}_j'
\right.
\nonumber \\
&+3\int_{t_{\rm{start}}}^t 
{\rm{d}}t'
\int {\rm{d}}^3\bm{\bar{x}}\;e^{-i{\bm{k}}\cdot \bm{\bar{x}}}
\xi(t',\bar{\bm{x}})
\rho_p(v_a(t',\bar{\bm{x}}),v_b(t',\bar{\bm{x}}))
\int {\rm{d}}^3\bm{x}'
e^{i{\bm{k}}\cdot \bm{x}'}
W
(x',t-t'
)
{V}_l(t',\bar{\bm{x}})\hat{\bm{x}}_i'\hat{\bm{x}}_j'\hat{\bm{x}}_l'
\nonumber \\
&
+
\left.6
\int_{t_{\rm{start}}}^t {\rm{d}}t'
\int {\rm{d}}^3\bm{\bar{x}}\;e^{-i{\bm{k}}\cdot \bm{\bar{x}}}
\xi(t',\bar{\bm{x}})
\rho_p(v_a(t',\bar{\bm{x}}),v_b(t',\bar{\bm{x}}))
\int {\rm{d}}^3\bm{x}'
e^{i{\bm{k}}\cdot \bm{x}'}
W
(x',t-t'
)
{V}_l(t',\bar{\bm{x}})
{V}_m(t',\bar{\bm{x}})\hat{\bm{x}}_i'\hat{\bm{x}}_j'\hat{\bm{x}}_l'\hat{\bm{x}}_m'
\right]
,
\label{General formula for energy momentum tensor2}
\end{align}
where we have treated $\bm{v}_a,\bm{v}_b$ and $\bm{V}$ as functions of the collision region coordinates $(t',\bar{\bm{x}})$. 
We find that the first and second terms do not contribute to the transverse-traceless (TT) component of $T_{ij}$, and only the third term provides a nonzero contribution, (see Appendix~\ref{append : Proof that TT vanish} for details). The two-point correlation function of $T_{ij}^{\rm{TT}}$ (whose contributions are shown schematically in Fig.~\Ref{fig : collision regions and the point where we evaluate stress tensor}) is given by 
\begin{align}
    (2\pi)^3\Pi(t_1,t_2,k)\delta^{(3)}(\bm{k}-\bm{q})
    \equiv
    \langle T_{ij}^{\rm{TT}}(t_1,\bm{k})T_{ij}^{*\rm{TT}}(t_2,\bm{q})\rangle,
\end{align}
where 
\begin{align}
    \Pi(t_1,t_2,k) 
    =
    \frac{36}{\beta^2 R_0^2l_{w0}^2}
    &\int_{-\infty}^{t_1}{\rm{d}}t_1'\int_{-\infty}^{t_2}{\rm{d}}t_2'
    \int_0^\infty {\rm{d}}r_1
\left[-\frac{3}{k^4r_1^4}\cos(kr_1)+\left(\frac{3}{k^5r_1^5}-\frac{1}{k^3r_1^3}\right)\sin(kr_1)\right]
\delta\left(r_1-(t_1-t_1')\right)
\nonumber \\
&\int_0^\infty {\rm{d}}r_2
\left[-\frac{3}{k^4r_2^4}\cos(kr_2)+\left(\frac{3}{k^5r_2^5}-\frac{1}{k^3r_2^3}\right)\sin(kr_2)\right]
\delta\left(r_2-(t_2-t_2')\right)
\int{\rm{d}}^3\bm{r}e^{-i\bm{k}\cdot\bm{r}}
U(t_1',t_2',\hat{\bm{k}},\bm{r}),
\label{eq : result of correlation function of stress tensor}
\end{align}
and 
\begin{align}
    U(t_1',t_2',\hat{\bm{k}},\bm{r})
    &\equiv 
    16\pi^2
    \Lambda_{ijkl}(\hat{\bm{k}})
    \langle
    S_{ij}(t_1',\bm 0)S_{kl}(t_2',\bm{r})\rangle\,.
    \label{eq : definition of correlation function S}
\end{align}
Here $S_{ij}(t',\bar {\bm x})\equiv \xi(t',\bar{\bm{x}})\rho_p(v_a(t',\bar{\bm{x}}),v_b(t',\bar{\bm{x}}))
V_i(t',\bar{\bm{x}})V_j(t',\bar{\bm{x}})$, and $\Lambda$ is the projection operator onto the TT mode, see Eq.\,(\ref{eq : definition of projection tensor}). The function $U$ is given by (see Appendix~\Ref{append : evaluation of vector correlation function})
\begin{align}
    U(t_1',t_2',\hat{\bm{k}},\bm{r})&=
\frac{\pi^4}{1152}
\Gamma_*^4
\kappa^2
\kappa_\rho^2
(\Delta V)^2
(l_{w0}R_0)^2
    \left(
    1-(\hat{\bm{k}}\cdot \hat{\bm{r}})^2\right)^2
    e^{-I(t_1',t_2',r)}
    e^{4(t_{\langle1,2\rangle}'-t_*)}
U_*(t_{1,2}',r),\;\;({\rm{with}}\;r>|t_{1,2}'|)\,,
\label{eq : explicit form of S}
\end{align}
where $t_{1,2}'\equiv t_1'-t_2'$, $t_{\langle 1,2\rangle}'\equiv (t_1'+t_2')/2$.  $U_*$ and $I$ are defined in Eqs.~\eqref{eq : definition of S*} and \eqref{eq : definition of I}, respectively. 
For the nucleation rate of bubbles per unit volume, we have used
\begin{align}
    \Gamma(t)=\Gamma_*e^{\beta(t-t_*)}.
    \label{eq : nucleation rate}
\end{align}
$\Gamma_*$ and $\beta$ are model-dependent parameters that can be calculated from the thermal potential of the scalar field in a given model. We have normalized $t$ such that $\beta=1$.

\section{Calculation of the Gravitational Wave Spectrum}
\label{sec : GW spectrum}

In this section, we calculate the GW spectrum produced by particles from bubble collision using the two-point correlation function of the stress-energy tensor calculated in the previous section. 

\subsection{Calculation}
The GW spectrum at production, using the Green function method as described in Ref.~\cite{Jinno:2017fby}, is (see Appendix\,\ref{sec : GW formalism} for further details)
\begin{align}
    \Omega_{\rm{GW}}^* &\equiv \frac{1}{\rho_{\rm{tot}}}
    \frac{{\rm{d}}\rho_{\rm{GW}}(t,k)}{{\rm{d}}\ln k}
    \nonumber \\
    &=
    \frac{2Gk^3}{\pi\rho_{\rm{tot}}}
    \int_{t_{\rm{start}}}^{t_{\rm{end}}}
    {\rm{d}}t_1
    \int_{t_{\rm{start}}}^{t_{\rm{end}}}
    {\rm{d}}t_2
    \cos(k(t_1-t_2))
    \Pi(t_1,t_2,k),
    \label{eq : Omega GW 1}
\end{align}
where $\rho_{\rm{tot}}$ is the total energy density of the Universe at the phase transition, and we have neglected cosmic expansion, assuming that the duration of phase transition is significantly shorter than a Hubble time. Here $t_{\rm{end}}$ denotes the  end of the phase transition. To solve the integrals numerically, we take $t_{\rm{start}}=-\infty$ and $t_{\rm{end}}=\infty$, as $|t_{\rm{start}}-t_*|\gg \beta^{-1}$ holds because we are assuming $\beta^{-1}\ll H^{-1}\sim t_*$ \cite{Jinno:2017fby}. Then we can artificially take $t_{\rm{start}}-t_*\to-\infty$, which implies $t_{\rm{start}}\to-\infty$. 
Using Eqs.~\eqref{eq : result of correlation function of stress tensor},~\eqref{eq : definition of correlation function S}, and~\eqref{eq : Omega GW 1}, and performing the $t_1$ and $t_2$ integrals, we obtain
\begin{align}
    \Omega_{\rm{GW}}^*
    =&\frac{\pi^3}{16}
    \kappa^2(\Delta V)^2\frac{Gk^3}{\rho_{\rm{tot}}}
    \Gamma_*^4
    \int_0^{\infty}{\rm{d}}r_1
    \int_0^{\infty}{\rm{d}}r_2
    \int_{-\infty}^{\infty}{\rm{d}}t_{1,2}'
    \int_{-\infty}^{\infty}{\rm{d}}t_{\langle 1,2\rangle}'
    \nonumber \\
    &\left[-\frac{3}{k^4r_1^4}\cos(kr_1)+\left(\frac{3}{k^5r_1^5}-\frac{1}{k^3r_1^3}\right)\sin(kr_1)\right]
    \left[-\frac{3}{k^4r_2^4}\cos(kr_2)+\left(\frac{3}{k^5r_2^5}-\frac{1}{k^3r_2^3}\right)\sin(kr_2)\right]
    \nonumber \\
    &
    \cos\left(
    kt_{1,2}'+
    k(r_1-r_2)
    \right)
\int{\rm{d}}^3\bm{r}e^{-i\bm{k}\cdot\bm{r}}
    e^{-I(t'_{1}, t'_{2},r)}
    e^{4(t_{\langle 1,2\rangle}'-t_*)}
    \left(
    1-(\hat{\bm{k}}\cdot \hat{\bm{r}})^2\right)^2
    U_*(t_{1,2}',r).
    \label{eq : Omega * GW 1}
\end{align}
To perform the $t_1$ and $t_2$ integrals, we have used 
\begin{align}
\int_{-\infty}^{\infty}{\rm{d}}t_1\int_{-\infty}^{\infty}{\rm{d}}t_2\int_{-\infty}^{t_1}{\rm{d}}t_1'\int_{-\infty}^{t_2}{\rm{d}}t_2'=\int_{-\infty}^{\infty}{\rm{d}}t_1'\int_{-\infty}^{\infty}{\rm{d}}t_2'\int_{t_1'}^{\infty}{\rm{d}}t_1\int_{t_2'}^{\infty}{\rm{d}}t_2.
\end{align}
Performing the $r_1$ and $r_2$ integrals using Eq.~\eqref{eq : formula for r1 and r2 integral}, we obtain
\begin{align}
    \Omega_{\rm{GW}}^*
    =&\frac{\pi^3}{2304}
\kappa^2(\Delta V)^2\frac{G}{\rho_{\rm{tot}}}
    \Gamma_*^4
    \times
    k
    \int_{-\infty}^{\infty}{\rm{d}}t_{1,2}'
    \int_{-\infty}^{\infty}{\rm{d}}t_{\langle 1,2\rangle}'
    \cos\left(
    kt_{1,2}'
    \right)
\int{\rm{d}}^3\bm{r}e^{i\bm{k}\cdot\bm{r}}
    e^{-I(t'_{1}, t'_{2},r)}
    e^{4(t_{\langle 1,2\rangle}'-t_*)}
    \left(
    1-(\hat{\bm{k}}\cdot \hat{\bm{r}})^2\right)^2
    U_*(t_{1,2}',r).
    \label{GW expression on the way 1}
\end{align}

Performing the $\Omega_{r}$ and $t_{\langle 1,2\rangle}'$ integrals using Eq.~\eqref{eq : formula for t12 integral} then gives 
\begin{align}
    \Omega_{\rm{GW}}^*=
    &
    \left(\frac{H}{\beta}\right)^2
    \left(\frac{\alpha}{1+\alpha}
    \right)^2
    \Delta_{\rm{GW}}^{pp}
    \left(
    \kappa,k/\beta
    \right),
    \label{eq : final result of GW spectrum}
\end{align}
where we have restored the $\beta$-dependence and defined 
\begin{align}
    \Delta_{\rm{GW}}^{pp}
    \left(
    \kappa,k
    \right)
    \equiv
    &
    \frac{\pi^3}{32}
    \kappa^2
    k^{-1}
    \int_0^{\infty}{\rm{d}}r
    \int_{-r}^{r}{\rm{d}}t_{1,2}'
    \mathcal{I}(t_{1,2}',r)^{-4}
    j_2(kr)
    \cos
    (
    k t_{1,2}'
    )
    U_*(t_{1,2}',r).
    \label{eq : Delta GW}
\end{align}
$\mathcal I$ is defined in Eq.~(\ref{eq : definition of calI}), and 
\begin{align}
    j_2(x)
    \equiv 
    -\frac{3}{x^2}\cos x +\left(\frac{3}{x^3}-\frac{1}{x}\right)\sin x
\end{align}
is the spherical Bessel function of the first kind. 

Meanwhile, the GW spectrum from the propagating, uncollided parts of the bubbles, the so-called envelope approximation for the scalar field or bubble wall contribution to GWs, is given by~\cite{Jinno:2022fom}
\begin{align}
    \Omega_{\rm{GW}}^{\rm{env}}=
    &
    \left(\frac{H}{\beta}\right)^2
    \left(\frac{\Delta V}{\rho_{\rm{tot}}}
    \right)^2
    \Delta_{\rm{GW}}^{\rm{env}}
    \left(
    k/\beta
    \right),
    \label{eq : GW spectrum from uncollided scalar walls}
\end{align}
where $\Delta_{\rm{GW}}^{\rm{env}}$ is defined in Ref.~\cite{Jinno_2017}.
In principle, we should add the energy density in these uncollided bubble walls to the energy density in particles when evaluating the stress-energy tensor and calculate the resulting overall GW spectrum. However, this makes the calculation extremely complicated, hence we avoid this and simply consider the GWs produced by the two sources in isolation. This misses the contribution from the interference term between the two sources in the two-point correlation function; however, since the uncollided regions only exist briefly (over the duration of the phase transition), whereas the particle distribution sources GWs over a longer period of time (Hubble time), the interference term is expected to provide a subdominant contribution to the overall spectrum.   

Finally, we redshift Eq.~\eqref{eq : final result of GW spectrum}, which corresponds to the GW spectrum at production, to the spectrum expected at present. The current frequency $f$ can be written in terms of the physical wavenumber $k$ at production as
\begin{align}
    f=\frac{k}{2\pi}
    \left(\frac{a_*}{a_0}\right)
=
2.63\times 10^{-6} 
{\rm{Hz}}
\times 
\left(
\frac{k}{\beta}
\right)
\left(
\frac{\beta}{H}
\right)
\left(
\frac{T}{100{\rm{GeV}}}
\right)
\left(
\frac{g_*(T)}{100}
\right)^{1/6}.
\end{align}
Likewise, since GWs redshift as radiation, the current GW energy density $\Omega_{\rm{GW}}$ can be written as (see e.g. Ref.~\cite{Inomata:2018epa}) 
\begin{align}
    \Omega_{\rm{GW}}h^2
    =
1.65\times 10^{-5}
    \left(\frac{H}{\beta}\right)^2
    \left(\frac{\alpha}{1+\alpha}
    \right)^2
    \left(
\frac{g_*(T)}{100}
\right)^{-1/3}
\Delta_{\rm{GW}}.
\label{eq:GWformula}
\end{align}

\subsection{Results and discussions}

We find that our results from the calculation above are robust for $k<\beta$, but run into some problematic behavior for $k\gtrsim\beta$. Numerically, the problem emerges from the irregular behavior of the two-point correlation of the source terms at short distances in configuration space (Eq.~(\ref{eq : definition of correlation function S}) in $r \to 0$ limit). 
Since two bubbles cannot be produced at the same position, the configuration-space correlation function of the source goes to zero in the short-distance limit.
This is in contrast to normal perturbation theory, where the configuration-space correlation function has a maximum in the short-distance limit. 
This configuration leads to a negative power spectrum in Fourier space at some frequencies in the $k \gtrsim \beta$ regime. Physically, this issue arises because in our present analysis, we only consider a configuration in which four distinct bubbles collide at two collision points, which simplifies the computation. However, the same bubble can also collide with two other bubbles at the two collisions points, and this is more likely to occur  in the short-distance, high-frequency regime; including this dominant process would likely fix the above problem, but this contribution requires more involved numerical computations. Since the regime of relevance for the GW signal of interest in this paper is $k<\beta$ (as will be explained below), we leave a detailed study of this contribution for future work.

\begin{figure}[t]
\centering
\includegraphics[width=1.0\linewidth]{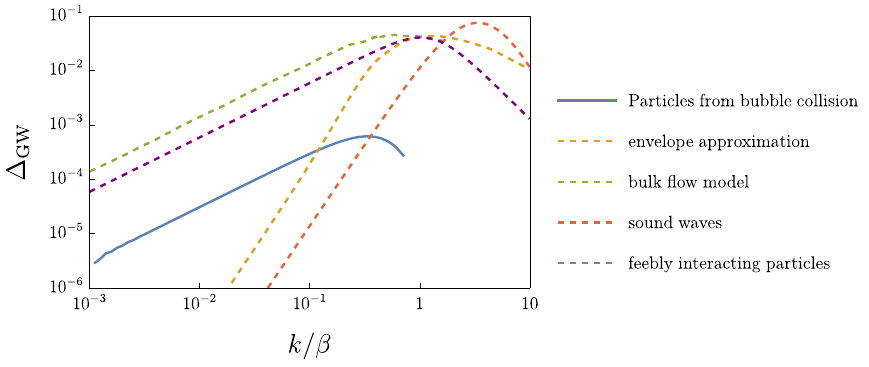} 
\label{fig : GW plot}
    \caption{Plot of $\Delta_{\rm{GW}}$ for various sources.  The solid blue curve corresponds to particles produced from bubble collisions, with $\kappa=1$. For comparison, we show $\Delta_{\rm{GW}}$ for the scalar field with the envelope approximation (orange dashed, from  Ref.~\cite{Jinno_2017}), the bulk-flow model (green dashed, from Ref.~\cite{Jinno:2017fby,Konstandin:2017sat}), sound waves (red dashed, made with {\tt{PTPlot}}\,\cite{Caprini:2019egz})) and shells of feebly-interacting particles (purple dashed, from Ref.~\cite{Jinno:2022fom}). 
    }
    \label{fig : GWspectrum}
\end{figure}

The GW spectrum obtained from the calculation described above is plotted in Fig.~\ref{fig : GWspectrum}. The blue curve corresponds to $\Delta_{\rm{GW}}^{pp}$ from particles produced from bubble collisions, Eq.~(\ref{eq : Delta GW}), with $\kappa=1$. This can be fit parametrically as 
\beq
\Delta_{\rm{GW}}^{{pp}~\text{(fit)}}(k/\beta)\,
\approx
\frac{0.003\,\kappa^2\, k/\beta}{1+16 (k/\beta)^{3}}\, ~~~~~ (\text{for }k/\beta\lesssim0.75).
\label{eq:fit1}
\eeq
hence we only plot $\Delta_{\rm{GW}}^{pp}$ up to $k/\beta=0.75$, for which we believe our results are reliable (as we discuss below, only the part below $k/\beta=0.1$ will be relevant for the overall signal). For comparison, we also plot, as dashed curves, the corresponding quantity $\Delta_{\rm{GW}}$ from other FOPT sources studied in the literature:
\begin{itemize}
    \item orange: scalar field or bubble wall contribution as calculated from the envelope approximation \cite{Jinno:2017fby}, which corresponds to the uncollided portions of bubble walls, 
    \item green: the bulk flow model \cite{Jinno:2017fby,Konstandin:2017sat}, with $\tau=100\,\beta^{-1}$, where $\tau$ is the dumping timescale of the energy density in the collided bubble walls \cite{Jinno:2017fby},
    \item red: sound waves, made with {\tt{PTPlot}}\,\cite{Caprini:2019egz}, 
    \item purple: shells of feebly-interacting particles \cite{Jinno:2022fom}.
    \end{itemize}

Physically, the GW signal from the scalar field (orange dashed curve), which corresponds to the signal sourced by the uncollided parts of the bubbles, always exists together with the GW signal from particles from bubble collisions. The plot shows that the latter has a smaller amplitude but a longer IR tail, and therefore is overwhelmed by the scalar field contribution at $k/\beta>0.1$ (where our numerical results are unstable), but dominates the signal at lower wave-numbers $k/\beta<0.1$, where our calculations are reliable. The other curves (bulk flow, sound waves, feebly interacting particles) are subdominant in runaway FOPTs where bubble collisions can efficiently produce particles, and are only shown for comparison. In particular, we see that the the spectrum for particles from bubble collisions has the same IR scaling $\propto k$ as the bulk flow (green) and feebly-interacting particles (purple) cases; this scaling is characteristic of sources that freely propagate without interacting long after the duration of the phase transition.

\begin{figure}[t]
\centering
\includegraphics[width=0.75\linewidth]{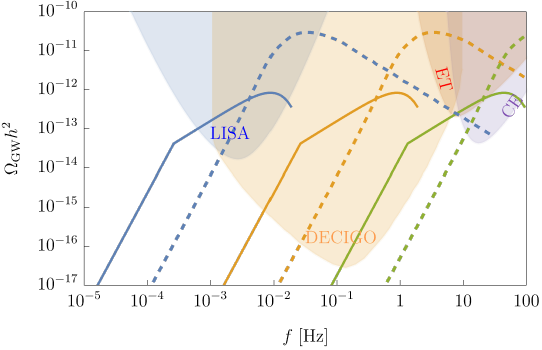}
    \caption{GW signals expected from FOPTs with runaway bubbles that efficiently produce particles from bubble collisions ($\kappa=1$), compared to power-law-integrated sensitivity curves for various upcoming GW detectors~\cite{LISA:2017pwj,Kawamura:2011zz,Punturo:2010zz,Reitze:2019iox} for 1 year observation time, with signal-to-noise ratio $=1$, obtained from \cite{Schmitz:2020syl}. The dashed and solid curves correspond to GWs produced from the scalar field with the envelope approximation and from the particles produced from bubble collisions, respectively. The three sets of curves correspond to phase transitions at temperatures $T=10,~1000,~5\times 10^4$ TeV (blue, orange, green), respectively. We fix $\beta/H=100$, $\alpha=10$, and $\kappa=1$.}
    \label{fig : GW plot and future sensitivity of GW detectors}
\end{figure}
Here, $\Delta_{\rm{GW}}^{pp}$ peaks at $k/\beta\sim 0.4$. This peak frequency is determined by the typical bubble size at collision, which represents the typical physical length scale over which the sources (particles) are correlated. 

The biggest difference between $\Delta_{\rm{GW}}^{pp}$ and the other curves in Fig.~\ref{fig : GWspectrum} is that the amplitude for $\Delta_{\rm{GW}}^{pp}$ is significantly smaller, by more than an order of magnitude. This is because the particles produced from bubble collision, although comparable in energy density to the other sources, are not as efficient at producing GWs, since GW production only occurs due to the quadrupolar anisotropy in the particle distribution in the cosmic or plasma frame.

In Fig.~\ref{fig : GW plot and future sensitivity of GW detectors}, we plot the GW signals expected from FOPTs at various temperatures with runaway bubbles that efficiently produce particles at collision ($\kappa=1$) against the power-law integrated sensitivities of various upcoming GW detectors. As discussed above, in such configurations, GWs are produced by two means: the scalar field or bubble wall contribution (dashed curves), and the particles produced from bubble collisions (solid curves), calculated in this work. We see that the latter component, while smaller in amplitude, has a longer IR tail, and therefore can dominate the GW signal at lower frequencies, modifying the overall signal in a manner that could be observable with these detectors. Moreover, the transition from $\propto f$ to $\propto f^3$ scaling where the two contributions become of comparable strength could be a distinguishing feature of such FOPT configurations. In the deep IR, beyond a Hubble time, we have artificially modified the IR tail to scale as $\propto f^3$ as dictated by causality~\cite{Caprini:2009fx, Cai:2019cdl}. This plot demonstrates that various upcoming space- or ground-based GW detectors will be sensitive to GWs from particles produced by bubble collisions at FOPTs across a broad range of energy scales.

\section{Summary}\label{sec:conclusion}
\label{sec : Discussions and conculsions}

In this paper, we have studied a new source of gravitational waves from first-order phase transitions: relativistic particles produced from the collisions of runaway bubbles. We demonstrated that such particles can account for a significant fraction of the vacuum energy released during the phase transition, and that the quadrupole anisotropy in their distribution in the cosmic or plasma frame can generate GWs. We also explained why other phenomenological models of energy flow after bubble collisions, such as the bulk-flow model, do not capture the physics of such configurations. This new component is relevant for any phase transition where bubble walls carry a significant fraction of the energy released during the transition, and where the background field has significant self-coupling or couplings to other fields, which would result in efficient particle production from bubble collisions.

We presented a semi-analytic calculation of the GWs produced by such particle distributions, which can be approximated with a simple, easy-to-use fit function, see Eqs.~(\ref{eq:GWformula},\ref{eq:fit1}). We found that this new contribution qualitatively modifies the overall GW signal from FOPTs when particle production from bubble collisions is an efficient process. While this new contribution has a smaller amplitude than the signal arising from the scalar field energy densities during collision, it can dominate the GW signal in the IR since the particles produced from bubble collisions persist long after the colliding bubbles have disappeared, producing GWs for a longer time at lower frequencies. This gives rise to a distinct transition of the spectral slope of the GW signal from a cubic to a linear falloff around $k/\beta \approx 0.1$ in the IR, which could provide a striking signature at upcoming GW detectors such as LISA, DECIGO, Cosmic Explorer, and the Einstein Telescope (see Fig.~\ref{fig : GW plot and future sensitivity of GW detectors}). 

Our semi-analytic approach made several simplifying approximations regarding the distribution of particles from bubble collisions and their subsequent evolution, which could be improved with future studies. In particular, numerical simulations of particle production from the bubble collision can better capture various details of the process and refine our results for the GW spectrum from this new contribution. Likewise, it would also be interesting to revisit the GW signal from various well-motivated BSM models that feature FOPTs in light of this new GW component. We leave these directions for future work. 

\section*{Acknowledgements}
We are grateful to Ryusuke Jinno and Thomas Konstandin for insightful discussions. KK was supported by the JSPS Overseas Challenge Program for Young Researchers and JST SPRING (grant number: JPMJSP2108). KI acknowledges the support of JSPS Postdoctoral Fellowship for Research Abroad.  This work was supported at Johns Hopkins by NSF Grant No.\ 2412361, NASA Grant No.\ 80NSSC24K1226, and the John Templeton Foundation.  BS is supported by the Deutsche Forschungsgemeinschaft under Germany’s Excellence Strategy - EXC 2121 “Quantum Universe” - 390833306. This work has been partially funded by the Deutsche Forschungsgemeinschaft (DFG, German Research Foundation) - 491245950. KK and BS thank the Johns Hopkins University for hospitality during the completion of this project.

\appendix

\section{Anisotropy of particle energy distribution in cosmic frame}
\label{append:boost}

In this appendix we derive the boost velocity Eq.\,\eqref{boostvelocity} and the anisotropic energy distribution function Eq.\,\eqref{boosted_distribution}, using the boost between reference frames as illustrated schematically in Fig.\,\ref{fig:boost}. 

In the cosmic frame, consider the collision of two bubble segments with velocities $\bm{v}_a$ and $\bm{v}_b$ at a collision point $C$, with coordinates $(t,\bm{x})$. The Lorentz boost factors of the two segments are $\gamma_a\equiv 1/\sqrt{1-v_a^2}$ and $\gamma_b\equiv 1/\sqrt{1-v_b^2}$. Let us denote the nucleation points of the two bubbles as $(t_{na},\bm{x}_{na})$ and $(t_{nb},\bm{x}_{nb})$, respectively. Then the coordinates of the collision region satisfy~\cite{Coleman:1977py}
\begin{align}
    -(t-t_{na})^2+(\bm{x}-\bm{x}_{na})^2=R_0^2,
    \nonumber \\
    -(t-t_{nb})^2+(\bm{x}-\bm{x}_{nb})^2=R_0^2,
\end{align}
where $R_0$ is a Lorentz-invariant constant determined by the thermal potential of the scalar field, and corresponds to the size of the bubble at nucleation. 
Taking the time derivatives of these equations yields~\cite{Coleman:1977py} 
\begin{align}
     \bm{x}-\bm{x}_{na}
    = R_0\gamma_a\hat{\bm{v}}_a,
    \;\;
    t-t_{na}
    \simeq |\bm{x}-\bm{x}_{na}|=R_0\gamma_a,
    \nonumber \\
    \bm{x}-\bm{x}_{nb}
    = R_0\gamma_b\hat{\bm{v}}_b,
    \;\;
    t-t_{nb}
    \simeq |\bm{x}-\bm{x}_{nb}|=R_0\gamma_b,
    \label{eq : relation between bubble radius and Lorentz boost factor}
\end{align}
where we used $\gamma_a\gg1$, $\gamma_b\gg 1$. 

We can Lorentz-boost to a frame $K$ (the center-of-collision frame) where the two bubbles have the same size, and the two wall segments have equal and opposite velocities (Fig.~\Ref{fig:boost}). Let us denote the velocity of the frame $K$ in the cosmic frame as $\bm{V}$, the coordinates of the collision point $C$ in frame $K$ as $(t',\bm{x}')$ and the nucleation points of the two bubbles as $(t_{na}',\bm{x}_{na}')$ and $(t_{nb}',\bm{x}_{nb}')$, respectively. We require that the bubbles are the same size in frame $K$, which means $
    \bm{x}'-\bm{x}_{na}' = -(\bm{x}'-\bm{x}_{nb}').$
Using
\begin{align}
   \bm{x}'-\bm{x}_{na(b)}' = \gamma(V) (\bm x - \bm x_{na(b)} - \bm V (t-t_{na(b)})),
\end{align}
where $\gamma(V)\equiv 1/\sqrt{1-V^2}$,
we obtain
\begin{equation} \bm{V}=\frac{\gamma_a\bm{v}_a+\gamma_b\bm{v}_b}{\gamma_a+\gamma_b}.
\end{equation}

As explained in the main text, we assume that the produced particles are isotropic in the frame $K$. 
Consider the injection of relativistic particles with energy ${\rm{d}}E(\hat{\bm{p}})$ and momentum ${\rm{d}}\bm{p} =\hat{\bm{p}} {\rm{d}}E(\hat{\bm{p}})$ in the momentum direction $\hat{\bm{p}}$. We denote the energy and the momentum of the corresponding particles in frame $K$ as ${\rm{d}}E'(\hat{\bm{p}}')$ and ${\rm{d}}\bm{p}'=\hat{\bm{p}}'{\rm{d}}E'(\hat{\bm{p}}')$, respectively. Assuming that the emitted particles are relativistic in both frames, these quantities are related as 
\begin{align}
   {\rm{d}}E(\hat{\bm{p}})
   =
   \gamma(V)
   ({\rm{d}}E'(\hat{\bm{p}}')+{\bm{V}}\cdot {\rm{d}}\bm{p}')
   =\gamma(V)
   (1+\bm{V}\cdot\hat{\bm{p}}'){\rm{d}}E'(\hat{\bm{p}}')
   =\frac{{\rm{d}}E'(\hat{\bm{p}}')}
   {\gamma(V)
   (1-\bm{V}\cdot \hat{\bm{p}})}.
   \label{eq : Lorentz transform of dE}
\end{align}
In the last equality, we used the Lorentz transformation of $\cos\theta\equiv \hat{\bm{p}}\cdot \bm{V}/V$ and $\cos\theta'\equiv \hat{\bm{p}}'\cdot \bm{V}/V$, given by
\begin{align}
   \cos\theta'=\frac{\cos\theta-V}{1-V\cos\theta}.
\end{align}
The differential solid angles ${\rm{d}}\Omega_{\bm{p}}$, ${\rm{d}}\Omega_{\bm{p}}'$ with respect to $\hat{\bm{p}}$, $\hat{\bm{p}}'$ are related as
\begin{align}
   {\rm{d}}{\Omega_{\bm{p}}}
   =
   \gamma(V)^2
   (1-\bm{V}\cdot \hat{\bm{p}})^2
   {\rm{d}}{\Omega_{\bm{p}}'}.
   \label{eq : Lorentz transform of Omega p}
\end{align}

Next, consider the total energy and the total momentum of the particles, $E$, $\bm{P}$ in the cosmic frame and $E', \bm{P}'(=0)$ in the center-of-collision frame. These are related as
\begin{align}
   E=\gamma(V)E'+\bm{V}\cdot\bm{P}'=\gamma(V)E'.
   \label{eq : Lorentz transform of E}
\end{align}
The volume of the collision region $\mathcal{V}$ in the cosmic frame and $\mathcal{V}'$ in the center-of-collision frame are related as
\begin{align}
   \mathcal{V}=\frac{1}{\gamma(V)}\mathcal{V}'.
   \label{eq : Lorentz transform of volume}
\end{align}
From Eqs.~\eqref{eq : Lorentz transform of E} and~\eqref{eq : Lorentz transform of volume}, the local energy densities of the produced particles ($\rho_p\equiv E/\mathcal{V}$) in the two frames are related as 
\begin{equation}
    \rho_p'=\frac{\rho_p}{\gamma(V)^2}.
    \label{eq : Lorentz transform of rho p}
\end{equation}
Likewise, using Eqs.~\eqref{eq : Lorentz transform of dE}, ~\eqref{eq : Lorentz transform of Omega p} and~\eqref{eq : Lorentz transform of rho p}, the angular distributions of the emitted particles in the two frames are related as
\begin{align}
   \frac{{\rm{d}}\rho_{p}}
    {{\rm{d}}{\Omega_{\bm{p}}}}(\hat{\bm{p}})
    =
    \frac{1}
    {\gamma(V)^2(1-\bm{V}\cdot \hat{\bm{p}})^3}
    \frac{{\rm{d}}\rho_{p}'}
    {{\rm{d}}{\Omega_{\bm{p}}'}}(\hat{\bm{p}}')
    =\frac{1}
    {4\pi\gamma(V)^4(1-\bm{V}\cdot \hat{\bm{p}})^3}
    \rho_p.
    \label{eq : d rhop d Omega p}
\end{align}
Here, we have assumed an isotropic distribution in the center-of-collision frame,
\begin{align}
    \frac{{\rm{d}}\rho_{p}'}
    {{\rm{d}}{\Omega_{\bm{p}}'}}(\hat{\bm{p}}')
    =
    \frac{1}{4\pi}\rho_p'.
\end{align}
Therefore, in the cosmic frame, the angular distribution of the energy density of emitted particles is given by
\begin{align}
   \frac{{\rm{d}}\rho_{p}}
    {{\rm{d}}{\Omega_{\bm{p}}}}(\hat{\bm{p}})
    &=
    \frac{1}
    {4\pi\gamma(V)^4(1-\bm{V}\cdot \hat{\bm{p}})^3}
    \rho_{p}.
\end{align}
Expanding this expression up to second order with respect to $\bm{V}\cdot {\hat{p}}$ 
and assuming $\gamma(V)\simeq 1$,\footnote{ $\gamma(V)\simeq 1$ holds as long as the two colliding bubbles sizes are of the same order. Cases where they differ by many orders of magnitude, although possible, are extremely rare events.} we obtain 
\begin{align}
   \frac{{\rm{d}}\rho_{p}}
    {{\rm{d}}{\Omega_{\bm{p}}}}(\hat{\bm{p}})
  &\simeq \frac{\rho_p}{4\pi}
    \left(
    1+3\bm{V}\cdot \hat{\bm{p}}+6(\bm{V}\cdot \hat{\bm{p}})^2
    \right).
\end{align}


\section{Transverse-traceless components of the stress-energy tensor in Eq.~\eqref{General formula for energy momentum tensor2}}
\label{append : Proof that TT vanish}

Here, we calculate the TT parts of the various terms in the stress-energy tensor as listed in Eq.~(\ref{General formula for energy momentum tensor2}).
For convenience, let us write Eq.~(\ref{General formula for energy momentum tensor2}) as 
\begin{align}
    T_{ij}(t,\bm k) = T_{ij,1}(t,\bm k) + T_{ij,2}(t,\bm k) + T_{ij,3}(t,\bm k),
\end{align}
where $T_{ij,a}$ denotes the $a$-th term in the expansion in Eq.~(\ref{General formula for energy momentum tensor2}).

The first term can be expressed as  
\begin{align}
T_{ij,1}(t,\bm{k})
&\propto\int^t {{\rm{d}}t'} 
\int {\rm{d}}^3\bm{\bar{x}}\;e^{-i{\bm{k}}\cdot \bm{\bar{x}}}
\xi(t',\bar{\bm{x}}) \rho_p(v_a(t',\bar{\bm{x}}),v_b(t',\bar{\bm{x}}))
\int {\rm{d}}^3\bm{x}'
e^{i{\bm{k}}\cdot \bm{x}'}
W
(x',t-t'
)
\hat{\bm{x}}_i'\hat{\bm{x}}_j'
\nonumber \\
&=
-\int^t {\rm{d}}t'
\int {\rm{d}}^3\bm{\bar{x}}\;e^{-i{\bm{k}}\cdot \bm{\bar{x}}}
\xi(t',\bar{\bm{x}})\rho_p(v_a(t',\bar{\bm{x}}),v_b(t',\bar{\bm{x}}))
\frac{\partial^2}{{\partial k_i}{\partial k_j}}
A(t-t',k).
\label{eq : energy momentum tensor 2}
\end{align}
Here, we defined 
\begin{equation}
    A(t-t',k)
    \equiv\int {\rm{d}}^3\bm{x}'
e^{i{\bm{k}}\cdot \bm{x}'}
\frac{1}{{x'}^2}
W
(x',t-t'
).
\end{equation}
Using the identity
\begin{equation}
    \frac{\partial^2}{{\partial k_i \partial k_j}}
A(t-t',k)
=\left(
-\frac{k_ik_j}{k^3}+\frac{\delta_{ij}}{k}
\right)
\frac{\partial A(t-t',k)}{\partial k}
+\frac{k_ik_j}{k^2}\frac{\partial^2A(t-t',k)}{\partial k^2},
\label{eq:dk2_a}
\end{equation}
we see that TT mode of Eq.~\eqref{eq : energy momentum tensor 2} vanishes because the terms proportional to $\delta_{ij}$ do not have traceless components and terms proportional to $k_i$ or $k_j$ do not have transverse components. 

The second term can be expressed as
\begin{align}
T_{ij,2}(t,\bm{k})
&\propto\int^t {{\rm{d}}t'} 
\int {\rm{d}}^3\bm{\bar{x}}\;e^{-i{\bm{k}}\cdot \bm{\bar{x}}}
\xi(t',\bar{\bm{x}}) \rho_p(v_a(t',\bar{\bm{x}}),v_b(t',\bar{\bm{x}}))
V_l(t',\bar{\bm{x}})
\int {\rm{d}}^3\bm{x}'
e^{i{\bm{k}}\cdot \bm{x}'}
W
(x',t-t'
)
\hat{\bm{x}}_l'\hat{\bm{x}}_i'\hat{\bm{x}}_j'
\nonumber \\
&=
i\int^t {\rm{d}}t'
\int {\rm{d}}^3\bm{\bar{x}}\;e^{-i{\bm{k}\cdot \bm{\bar{x}}}}
\xi(t',\bar{\bm{x}})\rho_p(v_a(t',\bar{\bm{x}}),v_b(t',\bar{\bm{x}}))
V_l(t',\bar{\bm{x}})
\frac{\partial^3}{{\partial k_i}{\partial k_j}{\partial k_l}}
B(t-t',k) \nonumber \\
&=i\int^t {\rm{d}}t'
\int {\rm{d}}^3\bm{\bar{x}}\;e^{-i{\bm{k}}\cdot \bm{\bar{x}}}
\xi(t',\bar{\bm{x}})\rho_p(v_a(t',\bar{\bm{x}}),v_b(t',\bar{\bm{x}}))
V_l(t',\bar{\bm{x}})
\frac{\partial^2}{{\partial k_i}{\partial k_j}}
\left(\hat{\bm{k}_l}\frac{\partial B(t-t',k)}{\partial k}\right),
\end{align}
with
\begin{align}
    B(t-t',k)
    \equiv\int {\rm{d}}^3\bm{x}'
e^{i{\bm{k}}\cdot \bm{x}'}
\frac{1}{{x'}^3}
W
(x',t-t').
\end{align}
We further calculate
\begin{align}
    &\frac{\partial^2}{{\partial k_i}{\partial k_j}}
\left(\hat{\bm{k}_l}\frac{\partial B({t-t',k})}{\partial {k}}\right)
\nonumber \\
&=\frac{\partial }{\partial k_i}\left[
\left(
-\frac{k_jk_l}{{k}^3}+\frac{\delta_{jl}}{{k}}
\right)
\frac{\partial B({t-t',k})}{\partial {k}}
+\frac{k_jk_l}{k^2}\frac{\partial^2B({t-t'},k)}{\partial k^2}
\right] \nonumber \\
&=\delta_{ij}\left(-\frac{{k}_l}{k^3}\frac{\partial B({t-t',k})}{\partial {k}}
+\frac{k_l}{{k}^2}\frac{\partial^2 B({t-t'},{k})}{\partial {k}^2}\right)
\nonumber \\
&\quad +
k_i\left[\left(\frac{3k_jk_l}{k^5}-\frac{\delta_{jl}}{k^3}\right)\frac{\partial B({t-t',k})}{\partial {k}}
+\left(-\frac{3k_jk_l}{{k}^4}+\frac{\delta_{jl}}{{k}^2}\right)\frac{\partial^2 B(t-t',k)}{\partial k^2}+\frac{k_jk_l}{{k}^3}
\frac{\partial^3 B({t-t',k})}{\partial {k}^3}\right] \nonumber \\
&\quad +k_j\left[-\frac{\delta_{il}}{k^3}\frac{\partial B({t-t',k})}{\partial k}
+\frac{\delta_{il}}{{k}^2}\frac{\partial^2 B({t-t',k})}{\partial {k}^2}\right].
\label{eq:dk3_b}
\end{align}
Again, terms that are proportional to $k_i$ or $k_j$ do not have transverse parts, while terms that are proportional to $\delta_{ij}$ do not have traceless parts. Therefore, $T_{ij,2}$ does not have any TT component. 

Finally, the third term is given by 
\begin{align}
    T_{ij,3}(t,\bm{k})
&=
 \frac{6}{\beta R_0l_{w0}}
\int_{t_{\rm{start}}}^t {\rm{d}}t'
\tilde{S}_{lm}(t',{\bm{k}})
\frac{\partial^4}{{\partial k_i}{\partial k_j}{\partial k_l}{\partial k_m}}
C(t-t',k),
\end{align}
where 
\begin{align}
    C(t-t',k)
    \equiv\int {\rm{d}}^3\bm{x}'
e^{i{\bm{k}}\cdot \bm{x}'}
\frac{1}{{x'}^4}
W\left(
x',t-t'
\right),
\label{eq : definition of C}
\end{align}
and
\begin{align}
    \tilde{S}_{lm}(t',{\bm{k}})\equiv
    \int {\rm{d}}^3\bm{\bar{x}}\;e^{-i{\bm{k}}\cdot \bm{\bar{x}}}
S_{lm}(t',\bar{\bm{x}}).
\end{align}
As mentioned in the main text, $S_{ij}(t',\bar {\bm x})\equiv \xi(t',\bar{\bm{x}})\rho_p(v_a(t',\bar{\bm{x}}),v_b(t',\bar{\bm{x}}))
V_i(t',\bar{\bm{x}})V_j(t',\bar{\bm{x}})$ is a stochastic variable depending on $(t',\bar{\bm{x}})$. For this term, we calculate
\begin{align}
\frac{\partial^4}{{\partial k_i}{\partial k_j}{\partial k_l}{\partial k_m}} 
C(t-t',k) 
&=(\delta_{im}\delta_{jl}+\delta_{il}\delta_{jm})
\left(
-\frac{1}{k^3}
\frac{\partial C(t-t',k)}{\partial k}
+\frac{1}{k^2}
\frac{\partial^2 C(t-t',k)}{\partial k^2}
\right)
\nonumber \\
&\quad +
\delta_{ij}\frac{\partial}{\partial k_m}
\left(-\frac{{\bm{k}}_l}{k^3}\frac{\partial C(t-t',k)}{\partial k}
+\frac{{\bm{k}}_l}{k^2}\frac{\partial^2 C(t-t',k)}{\partial k^2}\right)
\nonumber \\
&\quad + k_iD_{jkl}(\bm{k})+k_jE_{ikl}(\bm{k}),
\label{eq : differentiation of C with respect to k}
\end{align}
where $D_{jkl}$ and $E_{ikl}$ are tensor functions of $\bm{k}$, and we have made use of the expressions for the second and third derivatives in Eqs.~(\ref{eq:dk2_a}) and (\ref{eq:dk3_b}) to obtain this equation. Since terms proportional to $\delta_{ij}$ do not have traceless components and terms proportional to $k_i$ or $k_j$ do not have transverse components, the only relevant term for GW production is the first term. Thus the TT component of the stress-energy tensor is given by 
\begin{align}
    T_{ij}^{\rm{TT}}(t,\bm{k})
&=
 \frac{12}{\beta R_0l_{w0}}\int_{t_{\rm{start}}}^t {\rm{d}}t'
F(t-t',k)
\Lambda_{ijlm}(\bm{k})
\tilde{S}_{lm}(t',\bm{k}).
\label{eq : Deformation of Tij_TT}
\end{align}
Here, we have defined 
\begin{align}
    F(t-t',k)
    \equiv
    -\frac{1}{k^3}
\frac{\partial C(t-t',k)}{\partial k}
+\frac{1}{k^2}
\frac{\partial^2 C(t-t',k)}{\partial k^2}.
\label{eq : definition of F}
\end{align}
Using Eqs.~\eqref{eq : definition of W},~\eqref{eq : definition of C} and \eqref{eq : definition of F}, we obtain
\begin{align}
F(t-t',k)=
4\pi
\int_0^\infty {\rm{d}}r
\left[-\frac{3}{k^4r^4}\cos(kr)+\left(\frac{3}{k^5r^5}-\frac{1}{k^3r^3}\right)\sin(kr)\right]\delta\left(r-(t-t')\right).
\label{eq : F}
\end{align}
Using Eqs.~\eqref{eq : Deformation of Tij_TT} and~\eqref{eq : F} and taking the two-point correlation results in Eq.~\eqref{eq : result of correlation function of stress tensor}.


\section{Evaluation of two-point correlation function for the tensor field $S_{ij}$}
\label{append : evaluation of vector correlation function}

In this section, we calculate the correlation function $U$ defined in Eq.~\eqref{eq : definition of correlation function S}. 

\begin{figure}[t]
\centering
\includegraphics[width=1\linewidth]{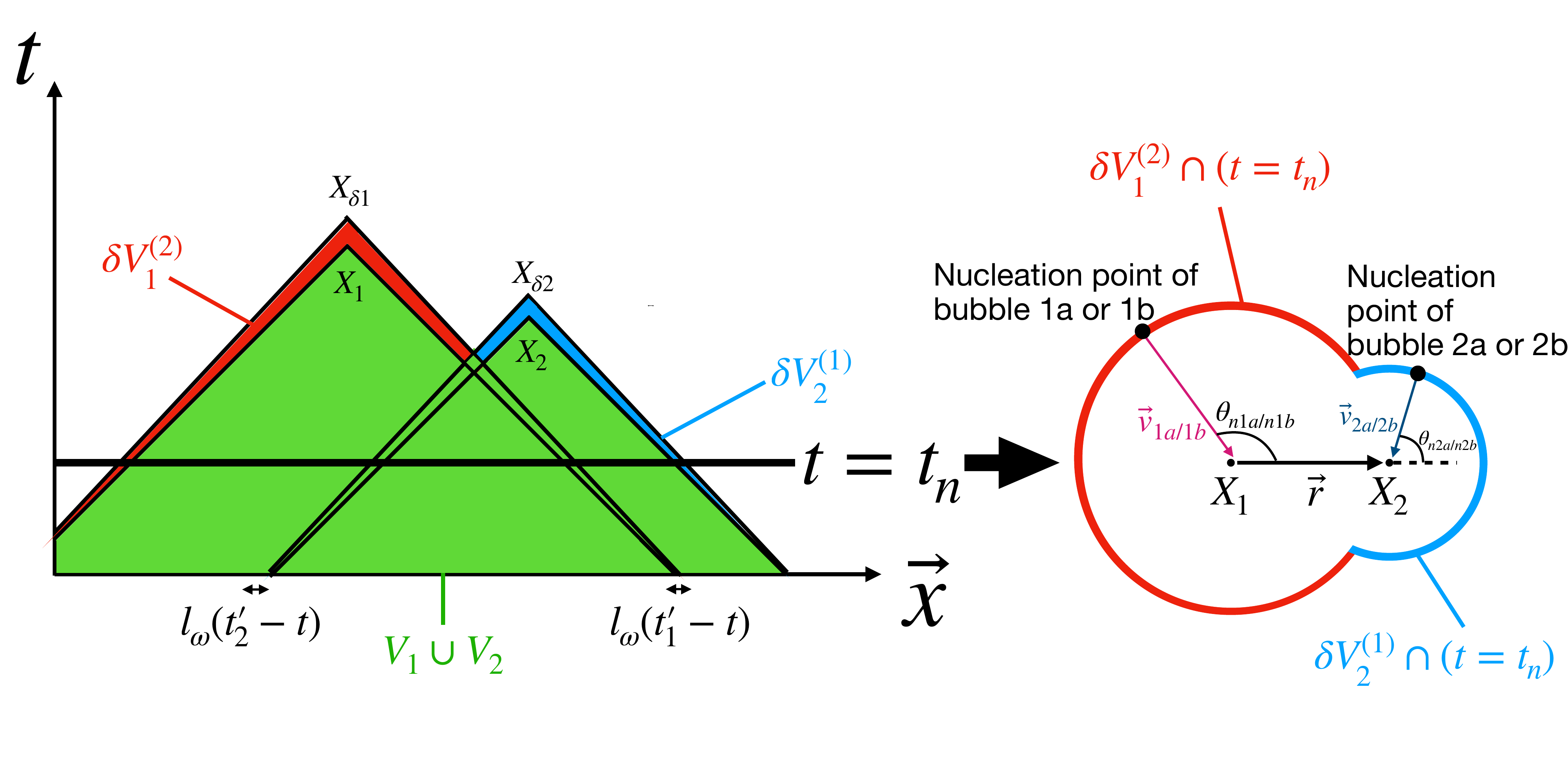}
    \caption{The schematic picture of the nucleation regions such that two bubbles collide at $X_1$ or $X_2$. The bubbles which collide at $X_1$ or $X_2$ should be nucleated in the red or blue region, respectively. We also prohibit bubble nucleation in the green region.
    }
    \label{fig : V1 and V2}
\end{figure}

Since we are interested in ultra-relativistic bubble walls, we approximate that all bubble walls propagate with velocity $v=1$. Consider past lightcones from points $X_1\equiv(t_1',\bm 0)$ and $X_2\equiv(t_2',\bm{r})$ (which serve as two different bubble collision sites), and denote these lightcones as $S_1$ and $S_2$. We denote the regions inside $S_1$, $S_2$ as $V_1$, $V_2$. We also define the points $X_{\delta_1}\equiv (t_1'+l_{w0},\bm 0)$, $X_{\delta_2}\equiv (t_2'+l_{w0},\bm{r})$
, and denote by $V_{\delta1}$ and $V_{\delta2}$ the regions inside the corresponding past lightcones modified with the radius of $t_{1(2)}'-t \to t_{1(2)}'-t+l_w(t_{1(2)}'-t)$, respectively.
This radius modification comes from the fact that the width of a bubble depends on time.
Next, we define 
$\delta V_1 \equiv V_{\delta1}-V_1$, $\delta V_2 \equiv V_{\delta2}-V_2$, $\delta V_1^{(2)}\equiv \delta V_{1}-(\delta V_1\cap V_{\delta2})$ (the red region in Fig.~\Ref{fig : V1 and V2}), and $\delta V_2^{(1)}\equiv \delta V_{2}-(\delta V_2\cap V_{\delta1})$ (the blue region in Fig.~\Ref{fig : V1 and V2}). Because we are focusing on the case where two bubbles pass the point $X_1$ and the other two bubbles pass the point $X_2$, the former two bubbles must have nucleated in the region $\delta V_1^{(2)}$ and the latter two bubbles in the region $\delta V_2^{(1)}$. We also require that the two points $X_1,X_2$ are in the false vacuum right before the collision, and thus forbid bubble nucleation in the region $V_1\cup V_2$ (the green region in Fig.~\Ref{fig : V1 and V2}). We focus on the case where the separation $r$ of the two points $X_1,X_2$ satisfies $r>|t_{1,2}'|$, where $t_{1,2}'\equiv t_1'-t_2'$, otherwise $S_1$ will lie within $S_2$, or vice versa. In this case, we cannot satisfy the conditions that $\bm{x}=\bm 0$ and $\bm{x}=\bm r$ are inside the false vacuum at $t< t_1'$ and $t<t_2'$, respectively.

The probability that bubbles do not nucleate in $V_1\cup V_2$ is given by $e^{-I(t_1',t_2',r)}$, where $I(t_1',t_2',r)$ is defined as~\cite{Jinno:2017fby}
\begin{align}
    I(t_1',t_2',r)=\beta^{-4}\Gamma_* e^{\beta(t_{\langle 1,2\rangle}'-t_*)}\mathcal{I}(t_{1,2}',r),
    \label{eq : definition of I}
\end{align}
with
\begin{align}
    \mathcal{I}(t_{1,2}',r)\equiv 8\pi \left(e^{\beta t_{1,2}'/2}+e^{-\beta t_{1,2}'/2}
    +\frac{\beta t_{1,2}^{'2}
    -\beta r^2-4r}
    {4r}
    e^{-\beta r/2}
    \right).
    \label{eq : definition of calI}
\end{align}

Now we define the regions where bubbles $1a,1b$, which collide with each other at region $X_1\equiv (t_1',\bm 0)$, are nucleated as $x_{n1a}=(t_{n1a},\bm{x}_{n1a})\in \delta V_1^{(2)}$ and $x_{n1b}=(t_{n1b},\bm{x}_{n1b})\in \delta V_1^{(2)}$, respectively.
We analogously define the regions where bubbles $2a,2b$, which collide with each other at $X_2\equiv (t_2',\bm{r})$, are nucleated as $x_{n2a}=(t_{n2a},\bm{x}_{n2a})\in \delta V_2^{(1)}$ and $x_{n2b}=(t_{n2b},\bm{x}_{n2b})\in \delta V_2^{(1)}$, respectively. 
Note that $x_{n1(2)a(b)}$ is the only exception for our notation $|\bm x| = x$.

Then, the correlation function of the tensor field $U$ defined in Eq.~\eqref{eq : definition of correlation function S} is given by
\begin{align}
    U(t_1',t_2',\hat{\bm{k}},\bm{r})
    &= 
    \frac{1}{4}e^{-I(t_1',t_2',r)}
    \int_{\delta V_1^{(2)}} {\rm{d}}^4x_{n1a} \Gamma(t_{n1a})
    \int_{\delta V_1^{(2)}} {\rm{d}}^4x_{n1b} \Gamma(t_{n1b})
    \int_{\delta V_2^{(1)}} {\rm{d}}^4x_{n2a} \Gamma(t_{n2a})
    \int_{\delta V_2^{(1)}} {\rm{d}}^4x_{n2b} \Gamma(t_{n2b})
    \nonumber \\
    &\times\Lambda_{ijkl}(\hat{\bm{k}})
    \rho_{p}(t_1',\bm 0)
    \rho_{p}(t_2',\bm{r})
    V_i(t_1',\bm 0)V_j(t_1',\bm 0)
    V_k(t_2',\bm{r})V_l(t_2',\bm{r}),
    \label{eq : definition of S}
\end{align}
where the particle energy densities $\rho$ are defined in terms of a fraction of the bubble wall energy densities,
\begin{align}
    \rho_{p}(t_1',\bm 0)\equiv \kappa \left(
    \rho_{\rm{wall}}(\bm{v}_{1a})
    +\rho_{\rm{wall}}(\bm{v}_{1b})
    \right), \nonumber \\
     \rho_{p}(t_2',\bm{r})\equiv \kappa \left(
    \rho_{\rm{wall}}(\bm{v}_{2a})
    +\rho_{\rm{wall}}(\bm{v}_{2b})
    \right),
    \label{eq : concrete expression of S}
\end{align}
and the boost velocities $\bm V$ are 
\begin{align}
    \bm{V}(t_1',\bm 0)
    \equiv
    \frac{\gamma_{1a}\hat{\bm{v}}_{1a}+\gamma_{1b}\hat{\bm{v}}_{1b}}{\gamma_{1a}+\gamma_{1b}},~~~~~
    \bm{V}(t_2',\bm{r})
    \equiv
    \frac{    \gamma_{2a}\hat{\bm{v}}_{2a}+\gamma_{2b}\hat{\bm{v}}_{2b}}{\gamma_{2a}+\gamma_{2b}}.
    \label{eq:Vs}
\end{align}
Here, we have used the subscripts $1a, 1b$ for the physical quantities related to the two bubbles colliding at $X_1$ and subscripts $2a,2b$ for the physical quantities related to the two bubbles colliding at $X_2$. The prefactor $1/4$ in Eq.~\eqref{eq : definition of S} accounts for the fact that these four bubbles cannot be distinguished from each other. 
For runaway bubbles, we approximate the Lorentz boost factors of each bubble $B$ by (see Eq.~(\ref{eq : relation between bubble radius and Lorentz boost factor})) 
\beq
\gamma_{B}=\frac{t_1'-t_{nB}}{R_0}.
\label{eq : approximations for Lorentz boost factors}
\eeq
Using Eqs.~\eqref{eq : definition of rho p},~\eqref{eq : definition of kappa},~\eqref{eq : approximations for Lorentz boost factors}, we can write
\begin{align}
    \rho_{p}(t_1',\bm 0)
    &=\frac{\kappa}{3}
    \frac{(t_1'-t_{n1a})^2+(t_1'-t_{n1b})^2}
    {R_0 l_{w0}}
    \Delta V,
    \nonumber \\
    \bm{V}(t_1',\bm 0)
    &=
    \frac{
    (t_1'-t_{n1a})\hat{\bm{v}}_{1a}
    +(t_1'-t_{n1b})\hat{\bm{v}}_{1b}
    }
    {(t_1'-t_{n1a})+(t_1'-t_{n1b})},
    \label{eq : concrete expressions for rho and V}
\end{align}
and the same relations hold for  $(t_2',\bm{r})$.

To proceed, we define $\bm x_{1a(b)} \equiv \bm x_{n1a(b)}$ and $\bm x_{2a(b)} \equiv \bm x_{n2a(b)} - \bm r$. 
We can express $\hat{\bm{x}}_{1a}$, $\hat{\bm{x}}_{1b}$, $\hat{\bm{x}}_{2a}$, $\hat{\bm{x}}_{2b}$, and $\hat{\bm{k}}$ with $\bm{r}$ parallel to $z$ axis:
\begin{align}
\hat{\bm{r}}&=(0,0,1)\nonumber \\
    \hat{{\bm{x}}}'_{1a}&=(\sin\theta_{n1a}\cos\varphi_{n1a},\sin\theta_{n1a}\sin\varphi_{n1a},\cos\theta_{n1a}), \nonumber \\
    \hat{{\bm{x}}}'_{1b}&=(\sin\theta_{n1b}\cos\varphi_{n1b},\sin\theta_{n1b}\sin\varphi_{n1b},\cos\theta_{n1b}), \nonumber \\
    \hat{{\bm{x}}}'_{2a}&=(\sin\theta_{n2a}\cos\varphi_{n2a},\sin\theta_{n2a}\sin\varphi_{n2a},\cos\theta_{n2a}), \nonumber \\
    \hat{{\bm{x}}}'_{2b}&=(\sin\theta_{n2b}\cos\varphi_{n2b},\sin\theta_{n2b}\sin\varphi_{n2b},\cos\theta_{n2b}), \nonumber \\
    \hat{\bm{k}}&=(\sin\theta_k\cos\varphi_k,\sin\theta_k\sin\varphi_k,\cos\theta_k).
\end{align}
Then the bubble wall velocities satisfy 
\begin{align}
    \hat{{\bm{v}}}'_{1a}&=
    -(\sin\theta_{n1a}\cos\varphi_{n1a},\sin\theta_{n1a}\sin\varphi_{n1a},\cos\theta_{n1a}), \nonumber \\
    \hat{{\bm{v}}}'_{1b}&=
    -(\sin\theta_{n1b}\cos\varphi_{n1b},\sin\theta_{n1b}\sin\varphi_{n1b},\cos\theta_{n1b}), \nonumber \\
    \hat{{\bm{v}}}'_{2a}&=
    -(\sin\theta_{n2a}\cos\varphi_{n2a},\sin\theta_{n2a}\sin\varphi_{n2a},\cos\theta_{n2a}), \nonumber \\
    \hat{{\bm{v}}}'_{2b}&=-(\sin\theta_{n2b}\cos\varphi_{n2b},\sin\theta_{n2b}\sin\varphi_{n2b},\cos\theta_{n2b}).
\end{align}

Using Eqs.~\eqref{eq : definition of S} and \eqref{eq:Vs}, we obtain 
\begin{align}
    U(t_1',t_2',\hat{\bm{k}},\bm{r})
    = 
    e^{-I(t_1',t_2',r)}&
    \int_{\delta V_1^{(2)}} {\rm{d}}^4x_{n1a} \Gamma(t_{n1a})
    \int_{\delta V_1^{(2)}} {\rm{d}}^4x_{n1b} \Gamma(t_{n1b})
    \int_{\delta V_2^{(1)}} {\rm{d}}^4x_{n2a} \Gamma(t_{n2a})
    \int_{\delta V_2^{(1)}} {\rm{d}}^4x_{n2b} \Gamma(t_{n2b})
    \nonumber \\
    &
    \times 
    \left(
    \gamma_{1a}+\gamma_{1b}
    \right)^{-2}
   \left(
    \gamma_{2a}+\gamma_{2b}
    \right)^{-2}
    \Lambda_{ijkl}(\hat{\bm{k}})
    \rho_{p}(t_1',\bm 0)
    \rho_{p}(t_2',\bm{r})
    \nonumber \\
    &\times \left(
\gamma_{1a}^2\gamma_{2a}^2\hat{v}_{1a,i}\hat{v}_{1a,j}\hat{v}_{2a,k}\hat{v}_{2a,l}
    \right.
    +\gamma_{1a}\gamma_{1b}
\gamma_{2a}\gamma_{2b}\hat{v}_{1a,i}\hat{v}_{1b,j}\hat{v}_{2a,k}\hat{v}_{2b,l}
    \nonumber \\
    &\qquad +\gamma_{1a}\gamma_{1b}\gamma_{2a}^2\hat{v}_{1a,i}\hat{v}_{1b,j}\hat{v}_{2a,k}\hat{v}_{2a,l}
    +\left.
\gamma_{1a}^2\gamma_{2a}\gamma_{2b}\hat{v}_{1a,i}\hat{v}_{1a,j}\hat{v}_{2a,k}\hat{v}_{2b,l}
    \right)
    ,
    \label{eq : deformation of S}
\end{align}
where we have used the symmetry under the exchange $a\leftrightarrow b$. 

We will now evaluate the various terms in the above equation one-by-one. 
~

\noindent \textbf{ First term in Eq.~\eqref{eq : deformation of S}:}

\begin{figure}[t]
\centering
\includegraphics[width=0.8\linewidth]{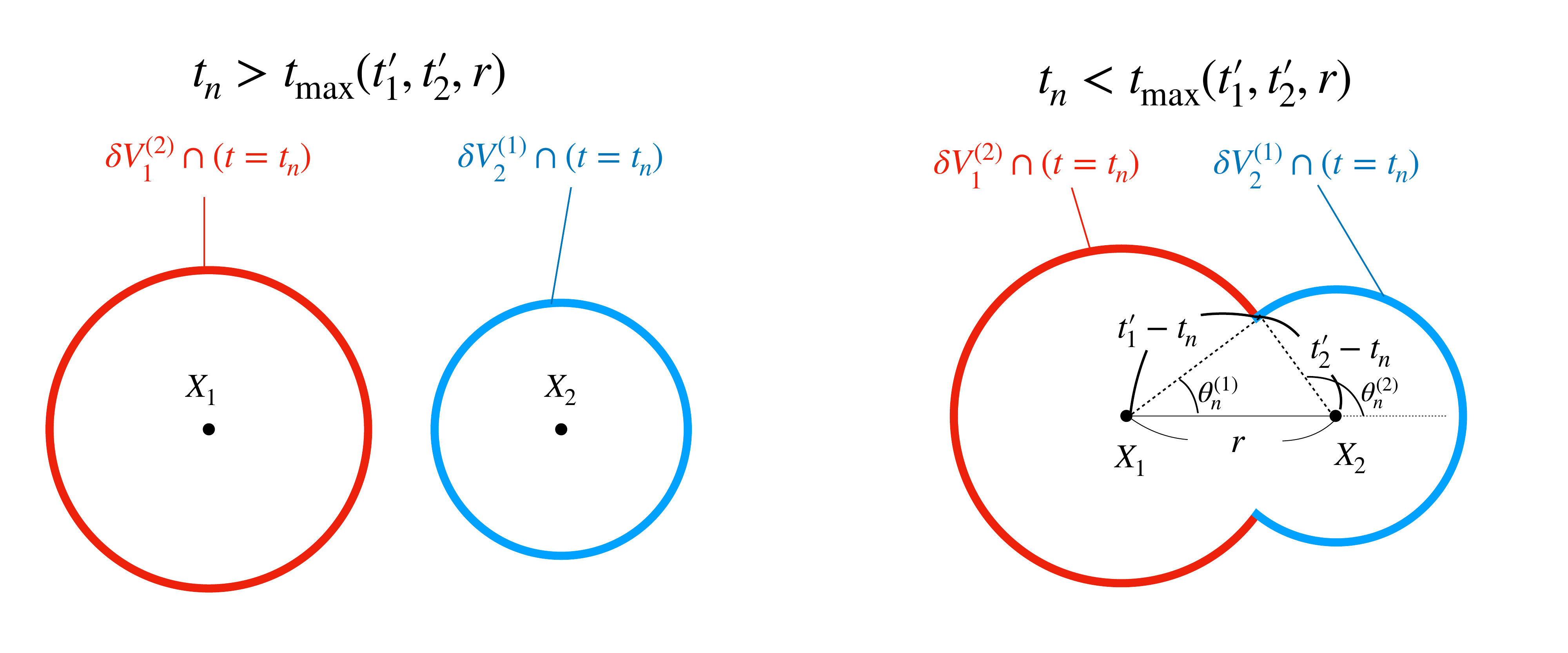}
    \caption{Schematic picture of the region $\delta V_1^{(2)}\cap (t=t_{n1a})$, $\delta V_2^{(1)}\cap(t=t_{n1a})$. The thickness of each regions are $l_w(t_1'-t_{n1a})$ and $l_w(t_2'-t_{n1a})$, respectively. 
    }
    \label{fig : integration region}
\end{figure}

For $\int_{\delta V_1^{(2)}}{\rm{d}}^4x_{n1a}$, we first perform the spatial integral over the region $\delta V_1^{(2)} \cap (t=t_{n1a})$ 
and then integrate over $t_{n1a}$. This region is a spherical shell with radius $r_B(t_1'-t_{n1a})\equiv t_1'-t_{n1a}$ and width $l_w(t_1'-t_{n1a})=l_{w0}R_0/(t_1'-t_{n1a})$. The spatial integral gives 
\begin{align}
\int{\rm{d}}^3\bm{x}\left|_{\delta V_1^{(2)}\cap (t=t_{n1a})}\right.= \int \sin\theta_{n1a}{\rm{d}}\theta_{n1a} \int{\rm{d}}\varphi_{n1a} r_B(t_1'-t_{n1a})^2l_w(t_1'-t_{n1a}). 
\label{eq : spacial integral}
\end{align}
The integration range of $\theta_{n1a}$ is shown in Fig.~\Ref{fig : integration region}. If $t_{n1a}>t_{\rm{max}}(t_1',t_2',r)\equiv t_{\langle 1,2\rangle}'-r/2$, where $t_{\langle 1,2\rangle}'\equiv (t_1'+t_2')/2$, then $\delta V_1^{(2)}\cap (t=t_{n1a})$ is a complete spherical shell. In this case, due to the spherical symmetry of the integral region, the integral over ${\rm{d}}\theta_{n1a}{\rm{d}}\varphi_{n1a}$ vanishes for the first term in Eq.~(\ref{eq : deformation of S}), which includes $v_{1a,i} v_{1a,j}$. On the other hand, if $t_{n1a}<t_{\rm{max}}(t_1',t_2',r)$, the integration range becomes $\theta_n^{(1)}(t_{n1a})<\theta_{n1a}<\pi$, where $\theta_n^{(1)}(t_{n1a})$ is defined by
\begin{align}
    \cos\theta_n^{(1)}(t_{n1})
    \equiv 
    \frac{r_B(t_1'-t_{n1})^2+r^2-r_B(t_2'-t_{n1})^2}{2r r_B(t_1'-t_{n1})}.
    \label{eq : cos theta n1}
\end{align}
In this case, the integral is non-vanishing, thus we focus on the region $t_{n1a}<t_{\rm{max}}$.

Similar arguments apply to $\int_{\delta V_2^{(1)}}{\rm{d}}^4x$. In this case, when $t_{n2a}<t_{\rm{max}}$, the integral range for ${\rm{d}}\theta_{n2a}$ becomes $0<\theta_{n2a}<\theta_n^{(2)}(t_{n2a})$, where $\theta_n^{(2)}(t_{n2})$ is defined by
\begin{align}
    \cos\theta_n^{(2)}(t_{n2})
    \equiv 
    -
    \frac{r_B(t_2'-t_{n2})^2+r^2-r_B(t_1'-t_{n2})^2}{2r r_B(t_2'-t_{n2})}.
    \label{eq : cos theta n2}
\end{align}
For $\int_{\delta V_1^{(2)}}{\rm{d}}^4x_{n1b}$ and $\int_{\delta V_1^{(2)}}{\rm{d}}^4x_{n2b}$, the integrals over $t_{\rm{max}}<t_{n1b}<t_1'$ and $t_{\rm{max}}<t_{n2b}<t_2'$ are non-vanishing because the integrand depends on neither $\hat{\bm{v}}_{1b}$ nor $\hat{\bm{v}}_{2b}$. 
Thus we obtain
\begin{align}
    U_1(t_{1,2}',r)\equiv&({\rm{The\;first\;term\;in\;Eq.~(\ref{eq : deformation of S})}})
    \nonumber \\
    = 
    e^{-I(t_1',t_2',r)}
    &\int_{-\infty}^{t_{\rm{max}}} {\rm{d}}t_{n1a} 
    r_B(t_1'-t_{1na})^2
    l_w(t_1'-t_{1na})
    \Gamma(t_{n1a})
    \int_{\theta_n^{(1)}(t_{n1a})}^{\pi}
    {\rm{d}}\theta_{n1a}
    \sin\theta_{n1a}
    \int_0^{2\pi}{\rm{d}}\varphi_{n1a}
    \nonumber \\
    &
    \times\int_{-\infty}^{t_{\rm{max}}} {\rm{d}}t_{n2a} 
    r_B(t_2'-t_{n2a})^2
    l_w(t_2'-t_{n2a})
    \Gamma(t_{n2a})
    \int_{0}^{\theta_n^{(2)}(t_{n2a})}
    {\rm{d}}\theta_{n2a}
    \sin\theta_{n2a}
    \int_0^{2\pi}{\rm{d}}\varphi_{n2a}
    \nonumber \\
    &\times\left[
    \int_{-\infty}^{t_{\rm{max}}} {\rm{d}}t_{n1b} 
    r_B(t_1'-t_{1nb})^2
    l_w(t_1'-t_{1nb})
    \Gamma(t_{n1b})
    \int_{\theta_n^{(1)}(t_{n1b})}^{\pi}
{\rm{d}}\theta_{n1b}
\sin\theta_{n1b}
\int_0^{2\pi}{\rm{d}}\varphi_{n1b}
\right.
\nonumber \\
&+
\left.
\int_{t_{\rm{max}}}^{t_1'} {\rm{d}}t_{n1b} 
    r_B(t_1'-t_{1nb})^2
    l_w(t_1'-t_{1nb})
    \Gamma(t_{n1b})
    \int_{0}^{\pi}
{\rm{d}}\theta_{n1b}
\sin\theta_{n1b}
\int_0^{2\pi}{\rm{d}}\varphi_{n1b}
\right]
    \nonumber \\
    &
    \times\left[
    \int_{-\infty}^{t_{\rm{max}}} {\rm{d}}t_{n2b} 
    r_B(t_2'-t_{n2b})^2
    l_w(t_2'-t_{n2b})
    \Gamma(t_{n2b})
    \int_{0}^{\theta_n^{(2)}(t_{n2b})}
    {\rm{d}}\theta_{n2b}
    \sin\theta_{n2b}
    \int_0^{2\pi}{\rm{d}}\varphi_{n2b}
    \right.
    \nonumber \\
    &
    +\left.
    \int_{t_{\rm{max}}}^{t_2'} {\rm{d}}t_{n2b} 
    r_B(t_2'-t_{n2b})^2
    l_w(t_2'-t_{n2b})
    \Gamma(t_{n2b})
    \int_{0}^{\pi}
    {\rm{d}}\theta_{n2b}
    \sin\theta_{n2b}
    \int_0^{2\pi}{\rm{d}}\varphi_{n2b}
    \right]
    \nonumber \\
    &
    \times
  \left(
    \gamma_{1a}+\gamma_{1b}
    \right)^{-2}
    \left(
    \gamma_{2a}+\gamma_{2b}
    \right)^{-2}
    \Lambda_{ijkl}(\hat{\bm{k}})
    \rho_{p}(t_1',\bm 0)
    \rho_{p}(t_2',\bm{r})
\gamma_{1a}^2\gamma_{2a}^2v_{1a,i}v_{1a,j}v_{2a,k}v_{2a,l}.
    \label{eq : deformation of the first term}
\end{align}

Using the expression of the projection operator (Eq.~\eqref{eq : definition of projection tensor}), we have
\begin{align}
    &\Lambda_{ijkl}(\hat{\bm{k}})
    v_{1a,i}v_{1a,j}
    v_{2a,k}v_{2a,l}
    \nonumber \\
    &
    =
    (\hat{\bm{v}}_{1a}\cdot
    \hat{\bm{v}}_{2a})^2
    -2(\hat{\bm{v}}_{1a}\cdot
    \hat{\bm{v}}_{2a})
    (\hat{\bm{v}}_{1a}\cdot
    \hat{\bm{k}})
    (\hat{\bm{v}}_{2a}\cdot
    \hat{\bm{k}})
    +\frac{1}{2}
    \left(
    (\hat{\bm{v}}_{1a}\cdot
    \hat{\bm{k}})^2
    (\hat{\bm{v}}_{2a}\cdot
    \hat{\bm{k}})^2
    -1
    +(\hat{\bm{v}}_{1a}\cdot
    \hat{\bm{k}})^2
    +(\hat{\bm{v}}_{2a}\cdot
    \hat{\bm{k}})^2
    \right),
    \label{eq : product of V multiplied by Lambda}
\end{align}
where we have used $v_a \simeq v_b \simeq 1$.
Integrating this over $\theta_{n1a}, \varphi_{n1a}, \theta_{n2a},$ and $\varphi_{n2a}$ gives
\begin{align}
    &\int_{\theta_n^{(1)}(t_{n1a})}^\pi {\rm{d}}\theta_{n1a}\sin\theta_{n1a}
    \int_0^{2\pi}{\rm{d}}\varphi_{n1a} 
    \int_{0}^{\theta_n^{(1)}(t_{n2a})}{\rm{d}}\theta_{n2a}\sin\theta_{n2a}
    \int_0^{2\pi}{\rm{d}}\varphi_{n2a}
    \left[\Lambda_{ijkl}(\hat{\bm{k}}) v_{1a,i}v_{1a,j} v_{2a,k}v_{2a,l} \right] \nonumber \\
    &= \frac{ \sin^4 \theta_k \pi^2 (r^2 - t_{1,2}^2)^2}{128 r^6 t_{n1a}'^3 t_{n2a}'^3}(r^2-t_{1,2}'(t_{1,2}'-2t_{n1a}'))
    (r^2-(t_{1,2}'-2t_{n1a}')^2) (r^2-t_{1,2}'(t_{1,2}'+2t_{n2a}'))
    (r^2-(t_{1,2}'+2t_{n2a}')^2). 
\end{align}

Using this and further performing the integrals over $\theta_{n1b},\varphi_{n1b}, \theta_{n2b},$ and $\varphi_{n2b}$, we can reexpress Eq.~\eqref{eq : deformation of the first term} as 
\begin{align}
U_1(t_{1,2}',r) =&
    \frac{\pi^4}{1152r^8}
    \kappa^2\kappa_\rho^2(\Delta V)^2
    (l_{w0}R_0)^2
    \sin^4\theta_k
    (r^2-t_{1,2}'^2)
    e^{-I(t_1',t_2',r)}
    \nonumber \\
    &
    \times\left[
    (r^2-t_{1,2}'^2)
    \int_{t_{\rm{min}}(t_{1,2}',r)}^{\infty}
    {\rm{d}}t_{n1a}'
    \int_{t_{\rm{min}}(t_{1,2}',r)}^{\infty}
    {\rm{d}}t_{n1b}'
    \right.
    \Gamma(t_1'-t_{n1a}')
    \Gamma(t_1'-t_{n1b}')
\frac{t_{n1a}'^2 + t_{n1b}'^2}{(t_{n1a}'+t_{n1b}')^2}
\nonumber \\
    &
    (r^2-t_{1,2}'(t_{1,2}'-2t_{n1a}'))
    (r^2-(t_{1,2}'-2t_{n1a}')^2)
    (r-t_{1,2}'+2t_{n1b}')
    \nonumber \\
    &
    +4r(r-t_{1,2}')
    \int_{t_{\rm{min}}(t_{1,2}',r)}^{\infty}
    {\rm{d}}t_{n1a}'
    \int_{0}^{t_{\rm{min}}(t_{1,2}',r)}
    {\rm{d}}t_{n1b}'
    \;
    \Gamma(t_1'-t_{n1a}')
    \Gamma(t_1'-t_{n1b}')
    t_{n1b}'\frac{t_{n1a}'^2 + t_{n1b}'^2}{(t_{n1a}'+t_{n1b}')^2}
    \nonumber \\
    &\left.(r^2-t_{1,2}'(t_{1,2}'-2t_{n1a}'))
    (r^2-(t_{1,2}'-2t_{n1a}')^2)
    \right]
    \nonumber \\
    \times&\left[
    (r^2-t_{1,2}^2)
    \int_{t_{\rm{min}}(-t_{1,2}',r)}^{\infty}
    {\rm{d}}t_{n2a}'
    \int_{t_{\rm{min}}(-t_{1,2}',r)}^{\infty}
    {\rm{d}}t_{n2b}'
    \right.
    \Gamma(t_2'-t_{n2a}')
    \Gamma(t_2'-t_{n2b}')
    \frac{t_{n2a}'^2 + t_{n2b}'^2}{(t_{n2a}'+t_{n2b}')^2}
\nonumber \\
    &
    (r^2-t_{1,2}'(t_{1,2}'+2t_{n2a}'))
    (r^2-(t_{1,2}'+2t_{n2a}')^2)
    (r+t_{1,2}'+2t_{n2b}')
    \nonumber \\
    &
    +4r(r+t_{1,2}')
    \int_{t_{\rm{min}}(-t_{1,2}',r)}^{\infty}
    {\rm{d}}t_{n2a}'
    \int_{0}^{t_{\rm{min}}(-t_{1,2}',r)}
    {\rm{d}}t_{n2b}'
    \;
    \Gamma(t_2'-t_{n2a}')
    \Gamma(t_2'-t_{n2b}')
    t_{n2b}'\frac{t_{n2a}'^2 + t_{n2b}'^2}{(t_{n2a}'+t_{n2b}')^2}
    \nonumber \\
    &\left.(r^2-t_{1,2}'(t_{1,2}'+2t_{n2a}'))
    (r^2-(t_{1,2}'+2t_{n2a}')^2)
    \right]
    .
    \label{eq:first_u}
\end{align}
Here, we have defined $t_{n1a}'\equiv t_1'-t_{n1a},\:t_{n1b}'\equiv t_1'-t_{n1b},\;t_{n2a}'\equiv t_2'-t_{n2a},\;t_{n2b}'\equiv t_2'-t_{n2b}$, and $t_{\rm{min}}(t_{1,2}',r)\equiv (t_{1,2}'+r)/2$, and made use of Eqs.~\eqref{eq : approximations for Lorentz boost factors} and \eqref{eq : concrete expressions for rho and V}. Note that the integration range for ${\rm{d}}t_{n1a}'$ is $t_{n1a}'>t_{\rm{min}}(t_{1,2}',r)$ because we are interested in the region $t_{n1a}<t_\text{max} $. Likewise, the integral range for ${\rm{d}}t_{n2a}'$ is $t_{n2a}'>t_{\rm{min}}(-t_{1,2}',r)=(r-t_{1,2}')/2$.
Using Eq.~\eqref{eq : nucleation rate}, the integration over $t_{n1a}',\;t_{n1b}',\;t_{n2a}',\;t_{n2b}'$ gives 
\begin{align}
 U_1(t_{1,2}',r)   =&
    \frac{\pi^4}{1152}
    \Gamma_*^4
    \kappa^2\kappa_\rho^2(\Delta V)^2
    (l_{w0}R_0)^2
    \sin^4\theta_k
    \frac{(r^2-t_{1,2}'^2)^2 }{r^8}
    e^{-I(t_1',t_2',r)}
    e^{4(t_{\langle 1,2\rangle}'-t_*)}
     U_{1,*}(t_{1,2}',r) U_{1,*}(-t_{1,2}',r)
    ,
    \label{eq : Result of the first term}
\end{align}
where 
\begin{align}
    &U_{1,*}(t_{1,2}',r)
    \equiv (r+t_{1,2}')
    \nonumber \\
    &\times \left[
    (r-t_{1,2}')\left((r^2-t_{1,2}'^2)^{2}J^{(1)}_{0,0}(t_{1,2}',r)
    +6t_{1,2}'(r^2-t_{1,2}'^2)J^{(1)}_{1,0}(t_{1,2}',r)
    \right.
    -4(r^2-3t_{1,2}'^2)J^{(1)}_{2,0}(t_{1,2}',r)
    -8t_{1,2}'J^{(1)}_{3,0}(t_{1,2}',r)
    \right)
    \nonumber \\
    &+\left.
    2(r^2-t_{1,2}'^2)^2J^{(1)}_{0,1}(t_{1,2}',r)
    +12t_{1,2}'(r^2-t_{1,2}'^2)J^{(1)}_{1,1}(t_{1,2}',r)
    -8(r^2-3t_{1,2}'^2)J^{(1)}_{2,1}(t_{1,2}',r)
    -16t_{1,2}'J^{(1)}_{3,1}(t_{1,2}',r)
    \right]
    \nonumber \\
    &+4r
    \left((r^2-t_{1,2}'^2)^2J_{0,1}^{(2)}(t_{1,2}',r)
    +6t_{1,2}'(r^2-t_{1,2}'^2)J_{1,1}^{(2)}(t_{1,2}',r)
    -4(r^2-3t_{1,2}'^2)J_{2,1}^{(2)}(t_{1,2}',r)-8t_{1,2}'J_{3,1}^{(2)}(t_{1,2}',r)
    \right).
    \label{eq : definition of S1*}
\end{align}
Here, $J^{(1)}_{m,n}(t_{1,2}',r)\equiv I^{(1)}_{m,n}(t_{\rm{min}}(t_{1,2}',r))$ and $J^{(2)}_{m,n}(t_{1,2}',r)\equiv I^{(2)}_{m,n}(t_{\rm{min}}(t_{1,2}',r))$, with
\begin{align}
    I^{(1)}_{m,n}(x)\equiv \int_x^{\infty}
    {\rm{d}}s
    \int_x^{\infty}
    {\rm{d}}t
    \frac{s^mt^n(s^2+t^2)}{(s+t)^2}e^{-s-t},
    \nonumber \\
    I^{(2)}_{m,n}(x)\equiv \int_x^{\infty}
    {\rm{d}}s
    \int_0^{x}
    {\rm{d}}t
    \frac{s^mt^n(s^2+t^2)}{(s+t)^2}e^{-s-t}.
    \label{eq : definition of integral function I}
\end{align}
Note that we have normalized $r,t$ such that $\beta=1$ in Eq.~\eqref{eq : nucleation rate}.

~

\noindent \textbf{Second term in Eq.~\eqref{eq : deformation of S}:}

This term gives
\begin{align}
    U(t_1',t_2',\hat{\bm{k}},\bm{r})
    = 
    e^{-I(t_1',t_2',r)}
    &\int_{-\infty}^{t_\text{max}} {\rm{d}}t_{n1a} 
    r_B(t_1'-t_{1na})^2
    l_w(t_1'-t_{1na})
    \Gamma(t_{n1a})
    \int_{\theta_n^{(1)}(t_{n1a})}^{\pi}
    {\rm{d}}\theta_{n1a}
    \sin\theta_{n1a}
    \int_0^{2\pi}{\rm{d}}\varphi_{n1a}
    \nonumber \\
    &\int_{-\infty}^{t_\text{max}} {\rm{d}}t_{n1b} 
    r_B(t_1'-t_{1nb})^2
    l_w(t_1'-t_{1nb})
    \Gamma(t_{n1b})
    \int_{\theta_n^{(1)}(t_{n1b})}^{\pi}
{\rm{d}}\theta_{n1b}
\sin\theta_{n1b}
\int_0^{2\pi}{\rm{d}}\varphi_{n1b}
    \nonumber \\
    &
    \int_{-\infty}^{t_\text{max}} {\rm{d}}t_{n2a} 
    r_B(t_2'-t_{n2a})^2
    l_w(t_2'-t_{n2a})
    \Gamma(t_{n2a})
    \int_{0}^{\theta_n^{(2)}(t_{n2a})}
    {\rm{d}}\theta_{n2a}
    \sin\theta_{n2a}
    \int_0^{2\pi}{\rm{d}}\varphi_{n2a}
    \nonumber \\
    &
    \int_{-\infty}^{t_\text{max}}{\rm{d}}t_{n2b} 
    r_B(t_2'-t_{n2b})^2
    l_w(t_2'-t_{n2b})
    \Gamma(t_{n2b})
    \int_{0}^{\theta_n^{(2)}(t_{n2b})}  
    {\rm{d}}\theta_{n2b}
    \sin\theta_{n2b}
    \int_0^{2\pi}{\rm{d}}\varphi_{n2b}
    \nonumber \\
    &
    \left(
    \gamma_{1a}+\gamma_{1b}
    \right)^{-2}
    \left(
    \gamma_{2a}+\gamma_{2b}
    \right)^{-2}
    \Lambda_{ijkl}(\hat{\bm{k}})
    \rho(t_1',\bm 0)
    \rho(t_2',\bm{r})
    \gamma_{1a}\gamma_{1b}
\gamma_{2a}\gamma_{2b}v_{1a,i}v_{1b,j}v_{2a,k}v_{2b,l}
    .
    \label{eq : deformation of the second term}
\end{align}
    
Using Eq.~\eqref{eq : definition of projection tensor}, we have
\begin{align}
    &\Lambda_{ijkl}(\hat{\bm{k}})
    v_{1a,i}v_{1b,j}
    v_{2a,k}v_{2b,l}
    \nonumber \\
    &
    =
    (\hat{\bm{v}}_{1a}\cdot
    \hat{\bm{v}}_{2a})
    (\hat{\bm{v}}_{1b}\cdot
    \hat{\bm{v}}_{2b})
    -(\hat{\bm{v}}_{1b}\cdot
    \hat{\bm{v}}_{2b})
    (\hat{\bm{v}}_{1a}\cdot
    \hat{\bm{k}})
    (\hat{\bm{v}}_{2a}\cdot
    \hat{\bm{k}})
    -(\hat{\bm{v}}_{1a}\cdot
    \hat{\bm{v}}_{2a})
    (\hat{\bm{v}}_{1b}\cdot
    \hat{\bm{k}})
    (\hat{\bm{v}}_{2b}\cdot
    \hat{\bm{k}})
    \nonumber \\
    &+\frac{1}{2}
    \left(
    (\hat{\bm{v}}_{1a}\cdot
    \hat{\bm{k}})
    (\hat{\bm{v}}_{1b}\cdot
    \hat{\bm{k}})
    (\hat{\bm{v}}_{2a}\cdot
    \hat{\bm{k}})
    (\hat{\bm{v}}_{2b}\cdot
    \hat{\bm{k}})
    -(\hat{\bm{v}}_{1a}\cdot
    \hat{\bm{v}}_{1b})
    (\hat{\bm{v}}_{2a}\cdot
    \hat{\bm{v}}_{2b})
    +(\hat{\bm{v}}_{1a}\cdot
    \hat{\bm{k}})
    (\hat{\bm{v}}_{1b}\cdot
    \hat{\bm{k}})
    (\hat{\bm{v}}_{2a}\cdot
    \hat{\bm{v}}_{2b})
    +(\hat{\bm{v}}_{1a}\cdot
    \hat{\bm{v}}_{1b})
    (\hat{\bm{v}}_{2a}\cdot\hat{\bm{k}})(\hat{\bm{v}}_{2b}\cdot\hat{\bm{k}})
    \right).
    \label{eq : product of V multiplied by Lambda}
\end{align}
Integrating this over the angles gives
\begin{align}
    &\int_{\theta_n^{(1)}(t_{n1a})}^\pi {\rm{d}}\theta_{n1a}\sin\theta_{n1a}
    \int_0^{2\pi}{\rm{d}}\varphi_{n1a} 
    \int_{0}^{\theta_n^{(1)}(t_{n2a})}{\rm{d}}\theta_{n2a}\sin\theta_{n2a}
    \int_0^{2\pi}{\rm{d}}\varphi_{n2a}\nonumber \\
    &\times\int_{\theta_n^{(1)}(t_{n1b})}^\pi {\rm{d}}\theta_{n1b}\sin\theta_{n1b}
    \int_0^{2\pi}{\rm{d}}\varphi_{n1b} 
    \int_{0}^{\theta_n^{(1)}(t_{n2b})}{\rm{d}}\theta_{n2b}\sin\theta_{n2b}
    \int_0^{2\pi}{\rm{d}}\varphi_{n2b}    
    \left[\Lambda_{ijkl}(\hat{\bm{k}})v_{1a,i}v_{1b,j}v_{2a,k}v_{2b,l}\right] \nonumber \\
    &= \frac{\sin^4 \theta_k \pi^4 (r^2 - t_{1,2}^2)^4}{512 r^8 t_{n1a}'^2 t_{n2a}'^2 t_{n1b}'^2 t_{n2b}'^2} (r^2-(t_{1,2}'-2t_{n1a}')^2) (r^2-(t_{1,2}'+2t_{n2a}')^2) (r^2-(t_{1,2}'-2t_{n1b}')^2) (r^2-(t_{1,2}'+2t_{n2b}')^2). 
\end{align}

Using this, we obtain
\begin{align}
&U_2(t_{1,2}',r)\equiv({\rm{The\;second\;term\;in\;Eq.~\eqref{eq : deformation of S}}})
    \nonumber \\
    &=
    \frac{\pi^4}{4608}
    \kappa^2\kappa_\rho^2(\Delta V)^2
    (l_{w0}R_0)^2
    \sin^4\theta_k
    \frac{(r^2-t_{1,2}'^2)^4}{r^8}
    e^{-I(t_1',t_2',r)}
    \int_{t_{\rm{min}}(t_{1,2}',r)}^{\infty}
    {\rm{d}}t_{n1a}'
    \int_{t_{\rm{min}}(t_{1,2}',r)}^{\infty}
    {\rm{d}}t_{n1b}'
    \int_{t_{\rm{min}}(-t_{1,2}',r)}^{\infty}
    {\rm{d}}t_{n2a}'
    \int_{t_{\rm{min}}(-t_{1,2}',r)}^{\infty}
    {\rm{d}}t_{n2b}'
    \nonumber \\
    &\quad
    \times\Gamma(t_1'-t_{n1a}')
    \Gamma(t_1'-t_{n1b}')
    \Gamma(t_2'-t_{n2a}')
    \Gamma(t_2'-t_{n2b}')
\frac{(t_{n1a}'^2+t_{n1b}'^2)(t_{n2a}'^2+t_{n2b}'^2)}{
(t_{n1a}'+t_{n1b}')^2(t_{n2a}'+t_{n2b}'
    )^2}
    \nonumber \\
    &\quad
    \times(r^2-(t_{1,2}'-2t_{n1a}')^2)
    (r^2-(t_{1,2}'-2t_{n1b}')^2)
    (r^2-(t_{1,2}'+2t_{n2a}')^2)
    (r^2-(t_{1,2}'+2t_{n2b}')^2)
    .
\end{align}
Here, we have used Eqs.~\eqref{eq : approximations for Lorentz boost factors} and~\eqref{eq : concrete expressions for rho and V}. 
Using Eq.~\eqref{eq : nucleation rate}, the integration over $t_{n1a}',\;t_{n1b}',\;t_{n2a}',\;t_{n2b}'$ gives
\begin{align}
U_2(t_{1,2}',r)=&
    \frac{\pi^4}{1152}
    \Gamma_*^4
    \kappa^2\kappa_\rho^2(\Delta V)^2
    (l_{w0}R_0)^2
    \sin^4\theta_k
    \frac{(r^2-t_{1,2}'^2)^2}{r^8}
    e^{-I(t_1',t_2',r)}
    e^{4(t_{\langle 1,2\rangle}'-t_*)}
    U_{2,*}(t_{1,2}',r)
    U_{2,*}(-t_{1,2}',r)
    ,
    \label{eq : Result of the second term}
\end{align}
where 
\begin{align}
    U_{2,*}(t_{1,2}',r) &
    \nonumber \\
    \equiv
    (r^2-t_{1,2}'^2)&
    \left[\frac{(r^2-t_{1,2}'^2)^2}{2}J_{0,0}^{(1)}(t_{1,2}',r)
    +2(r^2-t_{1,2}'^2)t_{1,2}'J_{1,0}^{(1)}(t_{1,2}',r)
    -2(r^2-t_{1,2}'^2)
    J_{2,0}^{(1)}(t_{1,2}',r)
    +2(r^2-t_{1,2}'^2)
    t_{1,2}'
    J_{0,1}^{(1)}(t_{1,2}',r)
    \right.
    \nonumber \\
    &
    \left.
    +8t_{1,2}'^2J_{1,1}^{(1)}(t_{1,2}',r)
    -16t_{1,2}'J_{2,1}^{(1)}(t_{1,2}',r)
    -2(r^2-t_{1,2}'^2)
    J_{2,0}^{(1)}(t_{1,2}',r)
    +8J_{2,2}^{(1)}(t_{1,2}',r)
    \right]
    .
    \label{eq : definition of S2*}
\end{align}

~

\noindent \textbf{Third and fourth terms in Eq.~\eqref{eq : deformation of S}:}

Similarly, the third and fourth terms give
\begin{align}
    \frac{\pi^4}{1152}
    \Gamma_*^4
    \kappa^2\kappa_\rho^2(\Delta V)^2
    (l_{w0}R_0)^2
    \sin^4\theta_k
    \frac{(r^2-t_{1,2}'^2)^2}{r^8}
    e^{-I(t_1',t_2',r)} 
    e^{4(t_{\langle 1,2\rangle}'-t_*)}
    \times \left(
    U_{1,*}(-t_{1,2}',r)
    U_{2,*}(t_{1,2}',r)
    +U_{1,*}(t_{1,2}',r)
    U_{2,*}(-t_{1,2}',r)
    \right)
    \label{eq : Result of the third and fourth term}
    .
\end{align}
~

\noindent \textbf{Final result for Eq.~\eqref{eq : deformation of S}:}

Combining Eqs.~\eqref{eq : Result of the first term},~\eqref{eq : Result of the second term} and~\eqref{eq : Result of the third and fourth term}, we obtain
\begin{align}
    U(t_1',t_2',\hat{\bm k},\bm r)
    =&
    \frac{\pi^4}{1152}
    \Gamma_*^4
    \kappa^2\kappa_\rho^2(\Delta V)^2
    (l_{w0}R_0)^2
    \sin^4\theta_k
    e^{-I(t_1',t_2',r)} 
    e^{4(t_{\langle 1,2\rangle}'-t_*)}
    U_*(t_{1,2}',r)
    ,
    \label{eq : final result of S}
\end{align}
where 
\begin{align}
    U_*(t_{1,2}',r)
    \equiv&
    \frac{(r^2-t_{1,2}'^2)^2}{r^8}
    \left(
    U_{1,*}(t_{1,2}',r)+
    U_{2,*}(t_{1,2}',r)
    \right)
    \left(
    U_{1,*}(-t_{1,2}',r)+
    U_{2,*}(-t_{1,2}',r)
    \right).
    \label{eq : definition of S*}
\end{align}

\section{GW Formalism}
\label{sec : GW formalism}

This appendix covers the basic definitions and relations for calculating the GW power spectrum in terms of the stress-energy tensor, based on Refs.~\cite{Jinno:2017fby, Jinno:2022fom}. We assume that the phase transition completes within a timescale significantly smaller than Hubble time, so that the expansion of space can be neglected. We start from Minkowski background with tensor perturbations as
\begin{align}
    {\rm{d}}s^2
    =
    -{\rm{d}}t^2
    +
    (\delta_{ij}+2h_{ij})
    {\rm{d}}x^i{\rm{d}}x^j.
\end{align}
The equation of motion of transverse and traceless tensor perturbations $h_{ij}$ is given by
\begin{align}
    \ddot{h}_{ij}(t,\bm{k})
    +
    k^2h_{ij}(t,\bm{k})
    =
    8\pi GT^{\rm{TT}}_{ij}(t,\bm{k}).
    \label{eq : EOM for tensor mode}
\end{align}
Here, $h_{ij}(t,\bm{k})$ is a Fourier component of $h_{ij}$, and we have defined
\begin{align}
    T^{\rm{TT}}_{ij}(\bm{k})
    \equiv
    \Lambda_{ijkl}(\hat{\bm{k}})
    T_{kl}(\bm{k}),
\end{align}
where the projection tensor $\Lambda_{ijkl}$ extracts the transverse and traceless (TT) part
\begin{equation}
    \Lambda_{ijkl}\equiv P_{ik}(\hat{\bm{k}})P_{jl}(\hat{\bm{k}})-\frac{1}{2}P_{ij}(\hat{\bm{k}})
    P_{kl}(\hat{\bm{k}}),
    \label{eq : definition of projection tensor}
\end{equation}
with $P_{ij}\equiv \delta_{ij}-\hat{\bm{k}}_i\hat{\bm{k}}_j$. Solving Eq.~\eqref{eq : EOM for tensor mode} with the Green function method gives
\begin{align}
    h_{ij}(t, \bm{k})
    =
    A_{ij}(\bm{k})\sin(k(t-t_{\rm{end}}))
    +B_{ij}(\bm{k})\cos(k(t-t_{\rm{end}}))
    \label{eq : solution of EOM for tensor mode}
\end{align}
with $t>t_{\rm{end}}$, where we assume the source $\Pi_{ij}$ is active from $t_{\rm{start}}$ to $t_{\rm{end}}$, and $A_{ij}$ and $B_{ij}$ are given by 
\begin{align}
    A_{ij}(\bm{k})
    \equiv
    \frac{8\pi G}{k}\int_{t_{\rm{start}}}^{t_{\rm{end}}}
    {\rm{d}}t'\cos(k(t_{\rm{end}}-t'))
    T^{\rm{TT}}_{ij}(t',\bm{k}),
    \label{eq : definition of Aij}
\end{align}
\begin{align}
    B_{ij}(\bm{k})
    \equiv
    \frac{8\pi G}{k}\int_{t_{\rm{start}}}^{t_{\rm{end}}}
    {\rm{d}}t'\sin(k(t_{\rm{end}}-t'))
    T^{\rm{TT}}_{ij}(t',\bm{k}).
    \label{eq : definition of Bij}
\end{align}
The GW power spectrum is defined as
\begin{align}
    \rho_{\rm{GW}}(t)
    \equiv
    \frac{\langle \dot{h}_{ij}(t,\bm{x})
    \dot{h}_{ij}(t,\bm{x})
    \rangle}
    {8\pi G},
\end{align}
where $\langle\dots\rangle$ denotes the oscillation average and also an ensemble average. We further define the GW density parameter $\Omega_{\rm{GW}}^*$ at production as
\begin{align}
    \Omega_{\rm{GW}}^* &\equiv \frac{1}{\rho_{\rm{tot}}}
    \frac{{\rm{d}}\rho_{\rm{GW}}(t,k)}{{\rm{d}}\ln k},
    \label{eq : definition of Omega GW1}
\end{align}
where $\rho_{\rm{tot}}$ is the total energy density at the time of the phase transition.

Using Eqs.~\eqref{eq : EOM for tensor mode},~\eqref{eq : definition of Aij},~\eqref{eq : definition of Bij} and~\eqref{eq : definition of Omega GW1}, we obtain
\begin{align}
    \Omega_{\rm{GW}}^*
    &=
    \frac{2Gk^3}{\pi\rho_{\rm{tot}}}
    \int_{t_{\rm{start}}}^{t_{\rm{end}}}
    {\rm{d}}t_1
    \int_{t_{\rm{start}}}^{t_{\rm{end}}}
    {\rm{d}}t_2
    \cos(k(t_1-t_2))
    \Pi(t_1,t_2,k),
    \label{eq : definition of Omega GW2}
\end{align}
where 
\begin{align}
    (2\pi)^3\Pi(t_1,t_2,k)\delta^{(3)}(\bm{k}-\bm{q})
    \equiv
    \langle T_{ij}^{\rm{TT}}(t_1,\bm{k})T_{ij}^{*\rm{TT}}(t_2,\bm{q})\rangle.
\end{align}

\section{Some integration formulae}
To derive Eq.~\eqref{eq : final result of GW spectrum}, we use the following formulae :
\begin{align}
    &\int_0^{\infty}{\rm{d}}x
    \sin x
    \left[-\frac{3}{x^4}\cos x +\left(\frac{3}{x^5}-\frac{1}{x^3}\right)\sin x\right]=
    \frac{1}{12},
    \nonumber \\
    &\int_0^{\infty}{\rm{d}}x
    \cos x
    \left[-\frac{3}{x^4}\cos x+\left(\frac{3}{x^5}-\frac{1}{x^3}\right)\sin x\right]=0,
    \label{eq : formula for r1 and r2 integral}
    \\
    &\int_{-\infty}^{\infty} {\rm{d}}x\exp\left(ax-be^{x}\right)=b^{-a}\Gamma(a) \;\;\;({\rm{with}}\; a>0,b>0).
    \label{eq : formula for t12 integral}
\end{align}
The functions $I_{m,n}^{(1)}(x)$ and $I_{m,n}^{(2)}(x)$ in Eq.~\eqref{eq : definition of integral function I} have the following analytic expressions:
\begin{align}
    I_{0,0}^{(1)}(x)
    &=
    -\frac{2}{3}
    e^{-2x}
    \left(
    -1+x+x^2+e^{2x}x^2(3+2x){\rm{Ei}}(-2x)
    \right),
    \nonumber \\
    I_{1,0}^{(1)}(x)
    &=
    \frac{1}{3}
    \left(
    e^{-2x}(2+x+x^2)+2x^3{\rm{Ei}}(-2x)
    \right),
    \nonumber \\
    I_{2,0}^{(1)}(x)
    &=
    \frac{1}{5}
    e^{-2x}
    \left(
    7-2x(-2+x(-2+x(2+x)))-2x^4e^{2x}(5+2x){\rm{Ei}}(-2x)
    \right),
    \nonumber \\
    I_{3,0}^{(1)}(x)
    &=
    \frac{1}{5}
    e^{-2x}
    \left(
    22+x(14+x(9+x+3x^2))+6e^{2x}x^5{\rm{Ei}}(-2x)
    \right),
    \nonumber \\
    I_{1,1}^{(1)}(x)
    &=
    \frac{1}{5}
    e^{-2x}
    \left(
    3+x(6+x+2x^2(2+x))+2e^{2x}x^4(5+2x){\rm{Ei}}(-2x)
    \right),
    \nonumber \\
    I_{2,1}^{(1)}(x)
    &=
    \frac{1}{5}
    e^{-2x}
    \left(
    6+x(12+x(7-(x-3)x))-2e^{2x}x^5{\rm{Ei}}(-2x)
    \right),
    \nonumber \\
    I_{3,1}^{(1)}(x)
    &=
    \frac{1}{7}
    e^{-2x}
    \left(
    26+x(52+x(31+x(16+x+2x^2(3+x))))+2e^{2x}x^6(7+2x){\rm{Ei}}(-2x)
    \right),
    \nonumber \\
    I_{2,2}^{(1)}(x)
    &=
\frac{2}{7}
    e^{-2x}
    \left(
    8+x(16-(x-2)x(8+x(7+x(x+5))))-e^{2x}x^6(7+2x){\rm{Ei}}(-2x)
    \right),
    \nonumber \\
    I_{0,1}^{(2)}(x)
    &=
    \frac{1}{6}
    e^{-2x}
    \left(
    -2(2+x+x^2)+e^x(4+x(2+x)(-1+3x))+e^{2x}x^3(-4{\rm{Ei}}(-2x)+(8+3x){\rm{Ei}}(-x))
    \right),
    \nonumber \\
    I_{1,1}^{(2)}(x)
    &=
    \frac{1}{5}
    e^{-2x}
    \left(
    -3-x(6+x+2x^2(2+x))+e^x(3+x(3+x(-1+x(3+2x))))+e^{2x}x^4(5+2x)(-2{\rm{Ei}}(-2x)+{\rm{Ei}}(-x))
    \right),
    \nonumber \\
    I_{2,1}^{(2)}(x)
    &=
    \frac{1}{15}
    e^{-2x}
    \left(
    3(-6+x(-12+x(-7+(x-3)x)))+e^x(18+x(18+x(9+x(x(7+5x)-2))))\right.
    \nonumber \\
    &\left.+e^{2x}x^5(6{\rm{Ei}}(-2x)+(12+5x){\rm{Ei}}(-x))
    \right),
    \nonumber \\
    I_{3,1}^{(2)}(x)
    &=
    \frac{1}{21}
    e^{-2x}
    \left(
    -3(26+x(52+x(31+x(16+x+2x^2(3+x)))))+e^x(78+x(78+x(39+x(13+2x(-1+x(4+3x))))))
    \right.
    \nonumber \\
    &\left.-2e^{2x}x^6(3(7+2x){\rm{Ei}}(-2x)-(7+3x){\rm{Ei}}(-x))
    \right),
    \label{eq : analytic formula for Modified version of I}
\end{align}
where we have defined 
\begin{equation}
{\rm{Ei}}(x)\equiv -\int_{-x}^{\infty}\dd t\, \frac{e^{-t}}{t}.
\end{equation}

\bibliography{Ref}

\end{document}